\documentclass[aps,pra,twocolumn,showpacs,floatfix]{revtex4-1}  

\newcommand{\eref}[1] {(\ref{#1})}
\newcommand{\Eref}[1] {Eq.~(\ref{#1})}
\newcommand{\Fref}[1] {Fig.~\ref{#1}}
\newcommand{\fref}[1] {Fig.~\ref{#1}}

\newcommand{\etal}{{\em et al.~}}

\usepackage{graphicx}
\usepackage{bm}        
\usepackage{amssymb}   
\usepackage{amsbsy}
\usepackage{amsmath} 
\usepackage{color}

\begin{document}

\title{Recent progress in the description of positron scattering from
  atoms using the Convergent Close-Coupling Theory}
\author{A~S~Kadyrov and I~Bray}
\affiliation{
Curtin Institute for Computation and Department of Physics, Astronomy and Medical Radiation Science,\\
Kent Street, Bentley, Perth, Western Australia 6102, Australia
}
\date{\today}

\begin{abstract}
Much progress in the theory of positron scattering on atoms has been
made in the ten years since the review of~\citet{SGB05}. We review
this progress for few-electron targets with a particular emphasis on
the two-centre convergent close-coupling and other theories which
explicitly treat positronium (Ps) formation. While substantial progress has
been made for Ps formation in positron scattering on few-electron targets,  considerable theoretical development is still required for multielectron atomic and molecular targets.
\end{abstract}

\maketitle
\tableofcontents

\section{Introduction}

There have been many reviews of positron physics over the
years~\cite{SGB05,GYS2010,MBR02,PositronPhysics,D98,Armour88,Charlton85,Ghosh1982,GH78}. 
More recently \citet{Laricchia08} and
\citet{CZ14} considered the subject with an emphasis on experimental
measurements involving noble gas targets. The related topic of
antihydrogen formation has also been thoroughly reviewed
\cite{Charlton2015,BBCM2015,HolzChNieto2004,charlton_antihydrogen_1994}.
Resonances and the closely related bound states of positrons with
atoms and molecules has also been extensively discussed~\cite{SGB05,0953-4075-40-2-F01,PhysRevA.80.032708,GYS2010,MBR02,ZC15}.

This work concentrates on the progress in application of theoretical
methods to scattering processes in a quantum few-body system
involving positrons as projectiles and multi-electron atomic targets
with explicit treatment of positronium formation.
Particular emphasis is on the developments taken place since the
comprehensive review of positron physics by \citet{SGB05}. It
begins by describing the currently available theories of 
 low-energy positron collisions with atoms and simple molecules. Then it
describes the development and application of the two-centre
convergent close-coupling method, which explicitly
treats the Ps-formation processes.

\subsection{Positron scattering}

Developments in positron physics have resulted in several technologies
in medicine and  material science. 
In medicine, the use of the positron emission tomography (PET)
scanners help to make  diagnoses of cancer detection and of certain
brain function disorders.
Material science uses  positron annihilation lifetime spectroscopy (PALS),
 to  analyze and design specific materials. The critical component is
 positronium (Ps) formation with its annihilation providing the key
 signature of its origin. Ps is a short-lived exotic atom of a bound
 positron-electron pair that has similar structure to atomic hydrogen. 

Scattering experiments are the main tool of modern physics to learn
about the structure of matter. By analyzing collision products we can
extract  useful information about the objects being studied. Historically,
ordinary matter particles like electrons and protons were
predominantly used as scattering particles in 
experimental atomic and molecular physics. With the development of positron and antiproton beams, studies of interactions of these particles with matter became possible. 

The last decade has seen significant progress in low-energy trap-based
positron beams \cite{SGB05,SJCMB08}. New high-resolution experimental measurements
have been performed for a range of atomic and molecular targets including He \cite{SMJCB08}, Ne and Ar
\cite{JCMSMMSB10,Jones_etal11}, Xe \cite{Machacek_etal11}, Kr \cite{Zetal11,Makochekanwa_etal11}
H$_2$ \cite{ZCSNB09} and H$_2$O 
\cite{ANU_H2O}. 
The development of positron beams  motivated novel
experimental and theoretical studies. Particularly important are the
positronium-formation cross sections, see the recent recommendations of \citet{Machacek16}.
In addition, interest has been motivated by possible binding of positrons to atoms~\cite{MBR02}.

Theoretical description of electron-impact ionisation (and excitation)
processes has
seen significant progress in recent years due to the development of 
various highly sophisticated methods including the exterior complex scaling (ECS) \cite{RBIM99,BRM01,BSB04l}, R-matrix with 
pseudo-states (RMPS) \cite{BHSBB96,B98cpc,Z2006cpc,ZB11l}, time-dependent 
close-coupling (TDCC) \cite{CP06} and convergent 
close coupling (CCC) \cite{BS92l,B02l,BFKS10}. A review of
electron-induced ionisation theory has been given by \citet{Bpr12}.
Such problems are examples of a class where
there is only one ``natural'' centre, namely the atomic centre. All
 coordinates are readily written
with the origin set at the atomic centre. Yet there are many
atomic collision systems of practical and scientific interest that
involve at least two centres, such as the positron-hydrogen scattering
system. 
This is a three-body system where all the particles are
distinguishable, and which allows for their rearrangement. Here we
have two ``natural'' 
centres, the atomic centre and the positronium (Ps) centre. For
positron-hydrogen scattering ionisation now splits into two separate
components: the rearrangement process of Ps-formation and the three-body
breakup process. A proper formulation of Ps-formation  
processes requires a combined basis consisting of two independent basis sets for each of the centres which
makes theoretical studies considerably more challenging than for
electron-scattering. Furthermore, the
positron-atom system is an ideal
prototype of the ubiquitous collision systems such as proton-atom
scattering, where charge-exchange processes also require a two-centre
treatment, see \citet{AKAB16} for example.

Every positron-atom scattering system
has an ionisation (breakup) threshold above which an electron may be
freely ejected. At 6.8~eV below this threshold is the Ps
formation threshold. At higher energies excitation
 or ionization of the target can take place. In addition, for
 multi-electron targets there could be many more reaction channels
 such as multiple 
 ionisation and ionisation-with-excitation. For molecular targets there
 could be rovibrational excitation and dissociation. Another reaction
 channel is the positron-electron
 pair annihilation, which  can occur at any scattering energy. In this
 process  the positron collides with one of the target's electrons and
 annihilates into 2 or 3 gamma rays. 
Theoretical \cite{RH98a,LW97} and experimental studies \cite{KGS96} of the annihilation process 
have shown that its cross section is up to 10$^{5}$ times smaller than
the elastic-scattering cross section.  Therefore the annihilation
channel is often omitted  from   
 scattering calculations. The elastic scattering, excitation, Ps
 formation and ionisation are the dominant channels of primary interest.

\subsubsection{atomic hydrogen}

The first theoretical studies of positron scattering from atoms date back to 
the 1950s, when \citet{MM54} used  the first Born approximation (FBA)
to describe Ps-formation in e$^+$-H collisions. The Born method is based on the assumption that the wavefunction for
the scattering system can be expanded in a rapidly convergent series.
This approximation consists in using plane waves to describe the
projectile and scattered particles.  The Born approximation is reliable
when the scattering potential is relatively small compared to
the incident energy, and thus is applicable only at high energies.
Therefore this method is mainly focused on high-energy excitation and ionization processes.

One of the most successful methods applied to the low-energy e$^+$-H elastic-scattering problem is 
the Kohn variational method.
The method was initially developed for scattering phase shifts 
  in nuclear reactions by \citet{Kohn48}.
 Later this method was extended to positron-hydrogen scattering by \citet{Schwartz61}.
 A detailed description of the method was given by \citet{AH91}.
The method is based on finding the form of the  functional (called
Kohn functional) involving the phaseshift as a 
parametric function of the total wavefunction. Requiring the functional to be stationary with
respect to the variations of the parameters, generates equations for
the linear parameters. The phaseshifts can be accurately obtained by    
 performing iterative calculations and finding the values of the
 nonlinear parameters that make the functional stationary.

The many-body theory of \citet{FW71} 
utilizes techniques that originated  from quantum
field theory. 
Using the Feynman-diagram  technique  the perturbation
series in the interaction between 
particles can be written in an  intuitive way. 
When it is applied to positron-atom scattering, however, 
 difficulties arise due to necessity to take into account
(virtual) Ps formation. However, a finite number of perturbation-theory terms cannot describe a bound Ps state.
\citet{GK94} developed a sophisticated method based on the many-body perturbation theory.
They used an approximation by considering virtual Ps formation only in the ground state. 
The calculations with this method showed  that for elastic scattering of positrons on hydrogen
(and helium atoms), the virtual Ps-formation contribution was almost 30\% and 20\%
 of the total correlation potential, respectively.
\citet{GL04} have further improved the method by introducing the techniques for the exact summation 
of the electron-positron ladder diagram series. The method  was applied to $\rm e^+-H$ 
scattering below the Ps-formation threshold, and resulted in good agreement with 
 accurate variational calculations.

The momentum-space coupled-channel optical (CCO) potential method 
  was first developed for electron-atom
 scattering in the 1980s by \citet{MS80,MS83}.
The method 
relies on constructing 
a complex equivalent local potential to account for the ionization and the Ps-formation channels.
  The CCO method gave excellent ionization \cite{Ratnavelu91}, total
  and Ps formation \cite{ZRM05} 
 cross sections for positron scattering on hydrogen. 
\citet{Ma_etal2011} used the CCO method to study excitation of atomic hydrogen from the
metastable 2s state. 

Following \citet{Macri04} \citet{JWZ2011} developed the so-called two-center two-channel eikonal final
state-continuum initial distorted-wave model to calculate Ps formation in the ground and the lowest excited states. 
They also presented a $n^{-s}$ scaling law for formation of Ps($n$) cross
sections on the entire energy range with $s$ varying as a
function of the positron incident energy.

A hyperspherical hidden crossing (HHC) method has been applied to
positron-impact ionization of hydrogen near the
threshold by \citet{JWSM2009}. They have calculated the ionization
cross-section for S, P and D-waves.
The HHC has also been used to calculate partial-wave
Ps(1$s$)-formation cross sections for low-energy positron collisions
with H, Li and Na atoms \cite{WS2012}.

One of the most sophisticated and commonly used methods is the close-coupling (CC) formalism, which 
is based on the expansion of the total wavefunction using the target-state
 wavefunctions. Substitution of this expansion into the Schr\"odinger equation 
yields coupled differential equations
in coordinate space, or
 Lippmann-Schwinger integral equations
for the T-matrix in momentum space. By solving these equations 
the transition amplitudes are obtained for all open channels.
Considerable pioneering work in this field has been done by
\citet{HNB90}, \citet{HB91}, \citet{M93} and \citet{Walters97}, 
who demonstrated the success of using two-centre expansions consisting
of Ps and atomic states. 

Here and below when we discuss close-coupling calculations we
denote  the combined two-centre  basis, used to expand the total
scattering wavefunction, as  ($N,M$), where $N$ is the number of
atomic (negative-energy) eigenstates and $M$ is the number of positronium 
eigenstates. We also use a bar to indicate (negative- and positive-energy)
pseudostates. For instance,
CC($\overline{N},M$) refers to close-coupling calculations with a
combined basis made of $N$ pseudostates for the atomic centre
supplemented by $M$ Ps eigenstates.

 \citet{HB91} performed the first accurate two-state CC(1,1)
 calculation. This work known as the static-exchange approximation
 showed a giant spurious resonance near 40~eV incident positron
 energy. Absence of such a resonance was demonstrated
 by  \citet{KMW95} using a larger CC($\overline{9},\overline{9}$)
 calculation that included $s$, $p$, and $d$-type pseudostates for
 both centres. However, the CC($\overline{9},\overline{9}$) gave new
 spurious resonances above the ionization 
threshold.  An energy-averaging procedure was used to get smooth results
for the cross sections to get rid of the pseudoresonances.

Considerable progress in the description of e$^+$-H has been made  by
\citet{M93} by using the close-coupling approach.
\citet{MR95} have performed convergence studies for the full
positron-hydrogen problem at low energies. Below the ionization
threshold they showed good agreement of sufficiently large
pseudostate close-coupling calculations and the benchmark variational
calculations of Ps formation by \citet{H84}.  

The convergent close-coupling (CCC) method was first developed for e$^-$-H
scattering by \citet{BS92}. Its modification to positron scattering in
a one-centre approach was trivial in that electron exchange was
dropped and the interaction potentials changed sign~\cite{BS93b,BS94}.
The CCC method with a 
($\overline{N},0$) basis, i.e. a single atomic-centre expansion
without any Ps states, gave very good results for the total,
elastic, excitation and ionization cross sections at higher incident
energies where the Ps-formation cross section is small allowing for
distinction between two experimental data sets~\cite{BS94}. The CCC calculations showed no pseudoresonances so long
as a sufficiently large basis was taken.  

Following the success of the large single-centre CCC calculations, \citet{KRMW96} and \citet{Mitroy96} 
used a large basis for the atomic centre supplemented by a few eigenstates of Ps.  
Calculations with the ($\overline{28},3$) and ($\overline{30},3$)
bases, made of a large atomic basis similar
to that of \citet{BS94} and the three lowest-lying
eigenstates of Ps, gave results significantly better than those from  the
($\overline{9},\overline{9}$) basis~\cite{KMW95}.

The CCC method with a  ($\overline{N},\overline{M}$) basis was developed by \citet{KB00jpbl} to study convergence in two-centre
expansions. This was applied to positron scattering on hydrogen within the S-wave model retaining only $s$-states in the combined basis. 
This work for the first time demonstrated the convergence of
the (non-orthogonal) two-centre expansions. The convergence in all channels was only possible when two independent near-complete Laguerre bases are employed on both of the centres. 
Interestingly, the total ionisation cross
section had two independently converged
components. One component was coming from the atomic centre and represented direct ionisation of the hydrogen atom, the other came from the Ps centre and represented Ps formation in the continuum. 
The convergence in the case of the full positron-hydrogen scattering problem
was demonstrated by \citet{KB02}. The CCC calculations with such a combined two-centre basis have shown very good agreement with the experimental measurements of \citet{ZKKS97} and \citet{ucl}. 

\subsubsection{helium}

Theoretical investigations of positron scattering from 
helium has an additional
challenge due to the complexity of the target structure.
In multi-electron targets two-electron excitation (or ionization with
excitation) channels are usually excluded. This is a good
approximation as the contribution of these channels is 
typically two orders of magnitude smaller than the corresponding one-electron excitation processes~\cite{Barna2004}.

First calculations of e$^+$-He scattering have been  performed by \citet{MM61} in the FBA.
They  used only the ground states for He and Ps and
 obtained cross sections for elastic scattering and Ps formation in its ground state.
Their study highlighted the importance of the Ps-channel coupling with the
elastic channel and thereby motivated further studies. 
Another extensive study based on the Born approximation was presented by \citet{MGS80},
to estimate Ps-formation cross section in arbitrary S-states.
From the FBA studies it became clear that 
more sophisticated approaches to the problem were required.

The distorted-wave Born approximation (DWBA) results are obtained by using distorted wavefunctions in 
first-order calculations. This method can give more accurate results
than the FBA down to lower energies.
Studies utilizing the DWBA by  \citet{PMS83,PMS87} were applied to the helium 
$2^{1}S$ and $2^1P$ excitations by positrons 
in the energy range from near the threshold up to 150~eV. 
Although the agreement with the experimental data was not very satisfactory,
 the method indicated  the importance of the inclusion of the polarization potential
 in the excitation channels at low energies. 
 The most systematic study of the ionisation process within the framework of DWBA was carried out 
by \citet{CFKMPRSS87,CMS96}. They used Coulomb and plane waves and also included exchange effects. 
They obtained good agreement with the experimental results 
of \citet{FKRS86,KBCP90} and \citet{MAL96} over the energy range from 
near-threshold to 500~eV. 
  However, the most important and difficult channel, Ps formation,  was not included in the early DWBA studies.    
                              
 \citet{SKT86} calculated the differential and total cross sections for the 
excitation of the helium $2^1S$ state using the second-order DWBA method.  
Another DWBA method including Ps channels has been  reported by \citet{SenMandal2009} 
for intermediate to high scattering energies.
They have calculated Ps-formation cross section and  achieved good agreement with available experimental data above 60~eV.
However, their results were not accurate for Ps formation below 60~eV. 
 Considering the fact that Ps formation starts
at 17.8~eV and reaches its maximum around 40 eV,
the applicability of this method is quite limited.

\citet{ACCS76} 
 applied the random-phase approximation (RPA), based on many-body theory \cite{FW71},
to positron-helium scattering at low  energies.
By using an approximate account of virtual
Ps formation they obtained good agreement with the elastic-scattering experimental data of
\citet{JP73} and \citet{CCGH72} at the lowest energies.
A similar  RPA method was  
used by  \citet{V90} to calculate positron impact excitation
of He into 2$^1$S and 2$^1$P states.
As mentioned above, \citet{GK94} developed a more sophisticated method based on the many-body perturbation theory.
The calculations with this method showed  that 
the contribution from virtual Ps formation was significant. 
Applications of the method to various atomic targets were reported by
\citet{DFGK96} and \citet{Gribakin96}.  

The Kohn variational method was first applied to positron-helium collisions by Humberston et al. \cite{H73,CH75,CH77,RH95}.  
 A comprehensive study of positron-helium scattering with the Kohn variational method was given by
 \citet{RH99}.  They obtained very accurate cross sections at low energies. However, agreement with the experimental 
results of \citet{MNKT85}  and  \citet{SKPSJ78} for Ps-formation cross
section was qualitative, with a similar energy dependence but with
almost 25\% difference in magnitude. Nevertheless, very good agreement was obtained for the total cross section below 
the Ps-formation threshold.

The CCO method, mentioned earlier, was applied to positron scattering on helium by \citet{CZ07}.   
They calculated the total and Ps-formation cross sections from the Ps-formation threshold to 500~eV. 
 The calculated results agreed well with the corresponding experimental data except for  the data of
 \citet{GH78} for the total cross section in the energy range from 50 to 100~eV.
\citet{MMRS77} applied a polarized-orbital approximation  
method to low-energy elastic positron-helium scattering and
obtained good agreement with the experimental results. 
Other calculations using optical potentials were presented  by \citet{TN02} and by
\citet{GM98} for slow-positron scattering from helium. Elastic-scattering cross sections of both reports
were in good agreement with experimental data. 
In general, the optical-potential methods proved to be useful for calculations 
of total cross sections. They
 are problematic, however,  when applied to more detailed cross sections like target excitation
and Ps formation in excited states.

\citet{SO88}  utilized the classical-trajectory Monte-Carlo (CTMC) technique to model positron scattering. The method is described
 fully for ion-atom collisions by \citet{AP66} and
 by  \citet{OS77}. Using this technique, \citet{SO88} calculated
 differential ionization cross section for positron-helium and also
 positron-krypton collisions. The main advantage of this method is
 that it can  
describe dynamic effects occurring in  collisions. For instance, the CTMC calculations showed that the
 probability of positron scattering to large  angles after ionising the target, may be comparable, 
or even much greater than, the probability of positronium formation. They suggested that the disagreement between theory and experiment above 60~eV  might be resolved by accounting for the flux in the experiments measuring positronium formation due to positrons  scattered to angles that allow them to escape confinement.

\citet{TBB05} also applied the CTMC method  to helium ionization
 by positron impact.  
They obtained good agreement with experimental data of \citet{FKRS86}.  
Results of CTMC reported by \citet{SO88}  overestimate the
recent experimental data by \citet{Caradonna_etal09}
for the  Ps-formation cross section below 60~eV.
This questions the  
applicability  of the classical trajectory approach to positron
scattering at low and intermediate energies.  
                               
A very comprehensive study of positron-helium scattering using the close-coupling method was carried out 
 by  \citet{CMKW98}.  They used two kinds of expansions, the  first one 
consisting of 24 helium eigen- and pseudostates and the lowest three
 Ps eigenstates, and the second one with
 only 30 helium eigen- and pseudo-states. The helium-target structure
 was modeled using a frozen-core approximation, which can produce good
 excited states, but a less accurate ground state. 
 The atomic pseudostates were constructed using a Slater basis. 
For the 27-state approximation, only results in the energy range above
 the positronium-formation threshold were given.  
Results for lower energies were unsatisfactory, and it was suggested
 that this might be due to the lack of convergence 
 from the use of an inaccurate helium ground-state wavefunction.  
   The total cross sections from both the 27- and 30-state approaches
 agreed well with the experimental results of  
\citet{SKPSJ78}, \citet{Ketal81} and \citet{MNKT85} for the energy
 range above the 
 threshold of positronium formation. For lower energies, qualitative
 agreement was obtained in terms  
of the shape and the reproduction of the Ramsauer-Townsend minimum near 2 eV, 
while the theoretical results were a factor of 2 larger than the
 experimental data. 
The Ps-formation cross section from the 27-state approximation was
 in good agreement  
with the experimental data of  \citet{MAL96} up to about 60~eV and
 with the data of  
\citet{FDC83} and \citet{DCBP86} up to 90~eV.  
Above 100~eV the calculations were much lower than the experimental data of 
\citet{DCBP86}  and \citet{FKRS86} while being closer to the data
of \citet{OMC93}.

 Another close-coupling calculation by \citet{HNB92,HNB90}, 
using a few helium and positronium states with a one-electron description of 
the helium atom, showed less satisfactory agreement with the experimental data. 
 Chaudhuri and Adhikari \cite{CA97,CA98,CA99}  performed calculations using only 5 helium and 
3 positronium states in the expansion. Their results for Ps formation agreed well with the experimental results 
by \citet{MAL96} at energies near the Ps threshold  and displayed a better agreement with the data of 
\citet{OMC93} at the  higher energies. However, the theoretical results were much lower than the 
experimental data at energies near the maximum of the cross section.

\citet{ITS96} applied the hyper-spherical CC approach to the problem. 
However, they also considered
  helium as a one-electron target, and thus the excitation and Ps-formation cross sections were multiplied by factor of 2.
 Satisfactory results were obtained for the total Ps-formation and
He(2$^1$S) and He(2$^1$P) excitation cross sections. The method was not able to describe low-energy scattering mainly
because a one-electron approach to helium is not realistic at low energies.

Despite the obvious advantages of the above-mentioned close-coupling  calculations
in handling many scattering channels simultaneously, none was 
 able to describe low-energy elastic scattering.  
 In addition, the presence of pseudo-resonances in cross sections below the ionization threshold
\cite{CMKW98,HNB92} indicated that  there was room for improvement.  
The use of the frozen-core He states also needed some attention as this
yielded an inaccurate result for the ground state of helium.

The first application of the single-centre CCC method to 
 positron scattering on helium was made by \citet{WBFS04jpbl,WBFS04}, using  very accurate helium wavefunctions obtained within the 
 multi-core approximation.
Very accurate elastic cross section was obtained below the
Ps-formation threshold by  
using orbitals with very high angular momenta. It was suggested that
 the necessity 
for  inclusion of very high angular-momentum orbitals was to mimic the 
 virtual Ps-formation  processes.
 The method also gave accurate results for medium to high-energy
scattering processes except that it was unable to explicitly yield the
 Ps-formation cross section. Interestingly, the method was not able to produce
a converged result at the Ore-gap region (where only the elastic and
 Ps-formation channels are open), which we shall discuss later in
 sec.~\ref{intcont}. 
Comparison of the frozen-core and the multi-core results showed that the frozen-core
wavefunctions lead to around 10\%  higher cross sections. 
\citet{WBFS04jpbl,WBFS04} demonstrated that single-centre expansions 
can give correct results below the Ps-formation threshold and at high energies where the probability of Ps formation is small, but for the full solution of the problem inclusion of the Ps centre into expansions was required.  

The initial application of the two-centre CCC approach to the problem
was within the frozen-core approximation~\cite{1742-6596-194-7-072009,UKFB10}. It was then
extended to a multi-core treatment~\cite{UKFBS10}. While generally
good agreement was found with experiment from low to high energies
certain approximations made need to be highlighted. The key
problematic channels are those of the type Ps-He$^+$. Electron
exchange between Ps and He$^+$ is neglected. Excitation of He$^+$ is
also neglected. While these may seem reasonable approximations for the
helium target due to its very high ionisation threshold (24.6~eV for
He and 54.4~eV for He$^+$),
they become more problematic for quasi two-electron targets such as
magnesium, discussed below.

Positron scattering from the helium $\rm 2^3S$ metastable state has been
theoretically studied for the first time by \citet{UKFBS10pra} at low and intermediate energies. 
Converged results for the total, Ps-formation and breakup cross sections
have been obtained with a high degree of convergence.
The obtained cross sections turned out to be significantly larger
than those for scattering from the helium ground state. 

\subsubsection{alkali metals}

Alkali atoms have an ionisation threshold that is lower than
6.8~eV. Consequently,
for positron scattering on alkalis the elastic and Ps-formation
channels are open at all incident-positron energies. For this reason,
theory has to treat appropriately the ``competition''  for the valence electron
between the two positively charged centers, the singly-charged ionic core and
the positron.

Positron scattering on the lithium target was investigated by \citet{WardETAL_1989_p-Li-Na-K,%
  SarkarETAL_1988_p-Li-Na,KhanETAL_1987_p-Li} using a one-center expansion. 
  However, convergence was poor due to the absence of Ps-formation channels \cite{McAlindenETAL_1997_p-Li}.
Two-center expansion was
employed by \citet{Guha-Ghosh_1981_p-Li},
\citet{Basu-Ghosh_1986_p-Li}, \citet{Abdel-Raouf_1988_p-Li-Na},
\citet{HewittETAL_1992_p-Li}, \citet{McAlindenETAL_1997_p-Li} and
\citet{LeETAL_2005_p-Li-Na}. As expected, these approaches gave better agreement with the
experiment. For positron-lithium case the 
two-center CCC
approach to positron collisions with lithium 
was reported by \citet{LKBS10}. This is the most comprehensive study of the
problem on an energy range spanning six orders of magnitude. While
convergence was clearly established, and agreement with
experiment for the total Ps-formation cross section~\cite{Setal2002} is
satisfactory, smaller experimental uncertainties would be helpful
to provide a more stringent test of the theory.

Positron scattering from atomic sodium has been intensively studied
for more than two decades. The first theoretical calculations relied
on simple two-center decomposition of the system wavefunction with
only the ground states of sodium and Ps atoms taken into
account \cite{Guha-Manda-p-Na-1980,Nahar-1987,%
  Abdel-Raouf_1988_p-Li-Na}. Then, \citet{HNB93}
conducted more complex close-coupling calculations, adding several
low-lying excited states for each positively-charged center. The
obtained results turned out to be in reasonably good agreement
with experimental data for both total
\cite{KwanETAL_1991_TCS-Na-K,Kauppila-Na-1994} and Ps-formation
\cite{Zhou-Ps-form-1994} cross sections.

Further enlargement of the number of channels in the close-coupling
calculations revealed that the theoretical Ps-formation cross sections
\cite{Mitroy-Ratn-1994,Ryzhikh-Mitroy_1997_JPB_p-Na,CMKW98}
deviated systematically from the experimental results for low impact
energies. The experiment showed that the Ps-formation cross section became larger
with decreasing energy while the most refined theoretical calculations
utilizing different methods of solution predicted consistently the
opposite.

To resolve the discrepancy \citet{Setal2002}
conducted the experimental study on Ps formation in positron
collisions with Li and Na atoms. This experiment confirmed the earlier
results of \citet{Zhou-Ps-form-1994}. The authors managed to extend
the impact-energy range down to 0.1 eV where the discrepancy between
the theory and experiment was even larger for sodium.
In striking contrast, for lithium the reasonable
agreement of the measured cross section with the theoretical
predictions was obtained with the use of the same methodology.
\citet{Ke-PRA-2004} applied the optical-potential approach. They found
that their theoretical cross section increases with the decrease in
the impact energies below 1~eV, but faster than the experimental
results. Unfortunately, this result was obtained with the use of some
approximations, whose validity were not analyzed. It would be
instructive if the same optical-potential approach was applied to the case
of positron scattering on lithium. 

\citet{LeETAL_2005_p-Li-Na} calculated
Ps-formation in positron-alkali collisions with the use
of the hyper-spherical close-coupling method. Their results support
the previous theoretical data
\cite{Mitroy-Ratn-1994,Ryzhikh-Mitroy_1997_JPB_p-Na,CMKW98}.
Large two-center CCC
calculations of positron scattering by atomic sodium were reported
by \citet{LKBS12}. Despite being the most comprehensive to date there
was no resolution of the discrepancy with experiment for Ps formation at low energies,
which we will highlight in Sec~\ref{alkali-results}.

While the lighter alkali atoms are well-modeled by a frozen-core
Hartree-Fock approximation, or even an equivalent local core
potential, the heavier ones become more problematic. With a reduced
ionisation threshold positron interaction with the core electrons,
either directly or via exchange of the valence and the core electrons,
becomes a more important component of the interaction. To the best of
our knowledge this has not been addressed to a demonstrable level of convergence
by any theory. Nevertheless, assuming that such problems are
more likely to be a problem at the higher energies,
\citet{Letal13} considered threshold behaviour of the elastic and Ps-cross
sections, and their convergence properties at near-zero energies for
Li, Na and K. This work confirmed the expected threshold law proposed
by \citet{wigner_behavior_1948}, but was unable to resolve the
discrepancy with the positron-sodium experiment at low energies.
Some earlier studies by \citet{HNB93} and \citet{MKW96} at
low to intermediate energies were performed at a time when
convergence was computationally impossible to establish.

\citet{CRZ2012} further developed the CCO method to study positron scattering on rubidium
 at intermediate and high energies. They calculated the Ps-formation and total cross sections. 
 Their total cross section results appear to overestimate the experiment.

Though outside the scope of this review we note that the complex
scaling method was recently used to study resonance phenomena in
positron scattering on sodium~\cite{UJ15} and potassium~\cite{UJ16}.
\subsubsection{magnesium}

Magnesium can be thought of as a quasi two-electron target with the
core electrons being treated by the self-consistent field Hartree-Fock
approach. 
Positron scattering on magnesium is particularly interesting due to
a large resonance in elastic scattering identified at low energies
by \citet{Mitroy08}. This was confirmed, though at a slightly
different energy, by the one-centre calculations
of \citet{SFB11} which were able to be taken to convergence in the
energy region where only elastic scattering was possible. Minor
structure differences are likely to be responsible for the small
variation in the position of the resonance. Two-centre CCC
calculations~\cite{Rav_etal12} also reproduced the resonance, but had
to make substantial approximations when treating the Ps-Mg$^+$
interaction. This is even more problematic than in the case of helium
discussed above 
since now we have a multi-electron Hartree-Fock core. Agreement with
experiment is somewhat variable, but there are substantial
experimental uncertainties, particularly in the Ps-formation cross
section. Other theoretical studies of positron-magnesium scattering include those by \citet{Campeanu98,Bromley98,Gribakin96,HNBJ96,CMKW98}.

\subsubsection{inert gases}

Inert gases heavier than helium represent a particular challenge for
theory, which is unfortunate because they are readily accessible
experimentally~\cite{ZC2015,CZ14}. Just the target structure is quite
complicated, but some good progress has been made in the case of
electron scattering by \citet{ZB04,ZB10}. For positron scattering once
Ps forms, the residual ion is of the open-shell type making full electron
exchange incorporation particularly problematic. The relatively high
ionisation thresholds for such targets mean that there is always a
substantial Ore gap where the Ps-formation cross section may be quite
large, but unable to be obtained in one-centre calculations which are
constrained to have only elastic scattering as the open
channel. Nevertheless, outside the extended Ore gap, formed by the
Ps-formation and the ionisation thresholds, one-centre CCC
calculations can 
yield convergent results in good agreement with
experiment~\cite{FB12_njp}. There are also first-order perturbative
calculations by \citet{CMS96,DFGK96,0953-4075-41-2-025201,PMS02} and
some based on close-coupling with convergence not fully established,
see \citet{McAlindenWalters92,GBW04}.

\citet{GLG2014} studied positron scattering and annihilation on noble-gas atoms using many-body theory  at energies below the Ps-formation threshold. They demonstrated that at low energies, the many-body
theory is capable of providing accurate results. 
\citet{FG2014} used an impulse approximation to describe Ps scattering on inert gases and provided quantitative theoretical explanation for the experimentally-observed 
similarity between the Ps and electron scattering for equal
projectile velocities \cite{Brawley2010}. According to \citet{FG2014} this happens due to the
relatively weak binding and diffuse nature of Ps, and the
fact that electrons scatter more strongly than positrons off
atomic targets.

\citet{PDM2013}
developed a model-potential approach to positron scattering on
noble-gas atoms based on an adiabatic method that treats the positron
as a light nucleus. The method was applied to calculate the elastic cross section below the Ps-formation threshold. 

\subsubsection{molecular hydrogen}

Positron collisions with molecular hydrogen have been studied  extensively by various experimental groups over the last 30 years 
 \cite{HDHKPSS82,CGHW83,FDC83,Griffith84,DCBP86,FKRS88, KBCP90, MLC93, JFKM95,ZLKKS97, SGMBBS02, KPB06, ZCSNB09,MAMBS2013}. 
Theoretical studies of this scattering system are challenging because of 
the complexities associated with the molecular structure and 
its non-spherical nature. 
Rearrangement processes add another degree of complexity to the problem. 
Until recently theoretical studies \cite{SM1970,BRD1979,RRS1980, DT90, AH91, BMG91,BBGD91, RKW04, Arretche2006, Arretche_Lima06,SL08, MS08, ZBFT11, TMMA12}  have been focused only at certain energy regions. In addition, there are few theoretical studies which include the Ps-formation channels explicitly.
The first calculations of Ps-formation cross section \cite{SM1970,BRD1979,RRS1980,BMG91} were obtained with the use of the first Born approximation.
\citet{BBGD91} used a coupled-static model, which  only included the ground states of H$_2$ and Ps.
This simple model was until recently the only coupled-channel
calculation available. 
Comprehensive review of the positron interactions with atoms and molecules has been given by \citet{SGB05}.

\citet{FFK15} have recently reported the total cross section for positron scattering
from the ground state of H$_2$ below the Ps-formation threshold using
density functional theory with a single-center
expansion. Their results are in good agreement  with recent single-centre CCC calculations of \citet{ZFB13r} below 1~eV. \citet{FFK15} have also performed analysis of experimental
and theoretical uncertainties using a modified effective
range theory (MERT). They concluded that a practically constant value of the total cross
section between 3~eV and the Ps-formation threshold is likely to be an effect of virtual Ps
formation.

The recent single-centre CCC calculations of positron scattering on
molecular hydrogen by \citet{ZFB13r} and antiproton collisions with
H$_2$ by \citet{AKFB13l,AKFAB14} have shown that the CCC formalism can
also be successfully applied to molecular targets. In order to obtain
explicit Ps-formation cross section a two-centre approach is
required, with the first attempt presented by \citet{Uetal15}. They
found some major challenges associated with the Ps-H$_2^+$
channel. Some severe approximations were required in order to manage
the non-spherical H$_2^+$ ion. Nevertheless, some good agreement with
experiment was found, see Sec.~\ref{H2-results}, but considerably more work is required.

\subsection{Antihydrogen formation}

A major application of positron-hydrogen scattering is to provide a
mechanism for antihydrogen formation. The idea is fairly simple, with
some accurate calculations performed quite some time
ago~\cite{Humberston87}. The basic idea is to time reverse the
Ps-formation process to hydrogen formation from Ps scattering on a
proton, and then to use the resultant cross sections for the case
where the proton is replaced by an antiproton, and hence forming
antihydrogen
\begin{equation}
e^++{\rm H}\leftrightarrow {\rm Ps}+p^+\equiv e^-+\bar{\rm H}\leftrightarrow {\rm Ps}+p^-.
\label{antiHform}
\end{equation}
The advantage of antihydrogen formation via this process is that it is
exothermic and so the cross section tends to infinity as the relative
energy goes to 
zero~\cite{wigner_behavior_1948}. This behaviour is enhanced in the
case of excited states with degenerate energy
levels~\cite{F74}. Antihydrogen formation is presently particularly
topical due to several groups (AEgIS~\cite{Doser12,Krasnicky14,Aegis2015},
GBAR~\cite{GBAR11,vdWerf14,Perez15}, ATRAP~\cite{Storry04},
ASACUSA~\cite{ASACUSA} and ALPHA~\cite{Amole15}) attempting
to make it in sufficient quantity in order to perform spectroscopic
and gravitational experiments. \citet{Ketal15l} provided CCC results for
Ps energy starting at $10^{-5}$~eV, which suffices for currently
experimentally accessible energies of around 25~meV.
Recent calculations of the cross sections for these processes will be
discussed in Sec.~\ref{antih}.

In this section we gave a general historical overview of various theoretical
developments related to positron scattering on atomic targets and the
H$_2$ molecule. In the next section we consider in some  detail
basic features of the coupled-channel formalism mainly in the context
of convergent close-coupling method and discuss the latest results.  

\section{Theory}

\subsection{Atomic hydrogen}

Here we describe basics of the close coupling approach based on the momentum-space integral equations. We consider the simplest case of scattering in a system of three particles: positron (to be
denoted $\alpha$), proton ($\beta$) and electron ($e$). Let us also
call $\alpha$ the pair of proton with electron, $\beta$ - positron
with electron and $e$ - positron with proton. We neglect spin-orbit interactions. In this case spacial and spin parts of the total three-body wavefunction separate.  The latter can be ignored as it has no effect on scattering observables.  
The spacial part of the total three-body
scattering wavefunction satisfies
\begin{eqnarray}
(H-E) \Phi = 0,
\label{SEH}
\end{eqnarray}
where
\begin{eqnarray}
H = H_0 + V_{\alpha} + V_{\beta} + V_{e} \equiv H_0 + V
\end{eqnarray}
is the full Hamiltonian, $H_0$ is the three-free-particle Hamiltonian, 
$V_{i}$ is the Coulomb interaction between particles of pair $i$
($i=\alpha,\beta, e$). The total Hamiltonian can also be expressed in the following way
\begin{eqnarray}
\label{htot2}
H=H_{\gamma}+\frac{q_{\gamma}^2}{2 M_{\gamma}}+\overline{V}_{\gamma},
\end{eqnarray}
where $H_{\gamma}$ is the Hamiltonian of the bound pair 
$\gamma$, 
${\bm q}_{\gamma}$ is the momentum of free particle $\gamma$ 
relative to c.m. of the bound pair,
$M_{\gamma}$  is the reduced mass of the two fragments 
and
$\overline{V}_{\gamma}=V-V_{\gamma}$ is the interaction
potential of the free particle with the bound system in channel $\gamma$ ($\gamma=\alpha, \beta$).

\begin{figure}[htb]
\begin{center}
\includegraphics[width=85mm,angle=0]{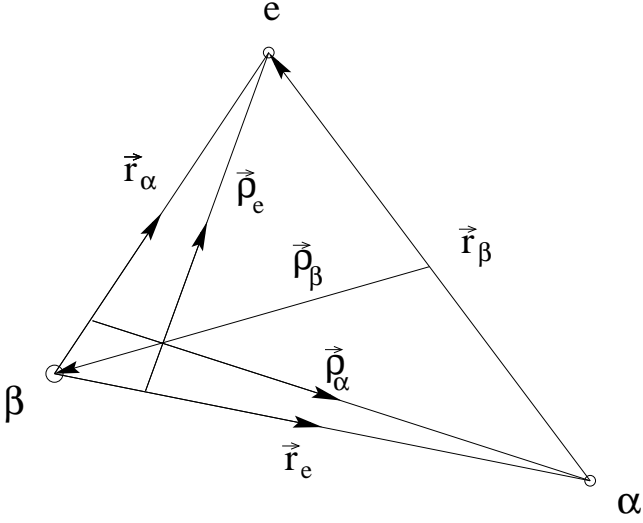} 
\caption{Jacobi coordinates for a system of three particles: positron
($\alpha$), proton ($\beta$), and electron ($e$).}
\label{Fig1Jacobi}
\end{center}
\end{figure}

Coupled-channel methods are based on expansion of the total wavefunction $\Phi$ in terms of functions of all
asymptotic channels. 
However, since the asymptotic wavefunction corresponding to 3 free particles has a complicated form \cite{MKP06,KMSBP03}, this is not practical. Therefore, we approximate $\Phi$ by expansion over some negative and discrete
positive energy pseudostates of pairs
$\alpha$ and $\beta$ especially chosen to best reproduce the corresponding
physical states. 
Suppose we have some $N_{\mathrm{H}}$ pseudostates in pair
$\alpha$ and $N_{\mathrm{Ps}}$ in pair $\beta$ satisfying the following conditions
\begin{eqnarray}
\langle \psi_{\alpha'}|\psi_{\alpha}\rangle=\delta_{\alpha'\alpha}, 
\quad
\langle \psi_{\alpha'}|H_{\alpha}|\psi_{\alpha}\rangle=\delta_{\alpha'\alpha}\varepsilon_{\alpha}
\label{diag1}
\end{eqnarray}
and
\begin{eqnarray}
\langle \psi_{\beta'}|\psi_{\beta}\rangle=\delta_{\beta'\beta}, \quad
\langle \psi_{\beta'}|H_{\beta}|\psi_{\beta}\rangle=\delta_{\beta'\beta}\varepsilon_{\beta},
\label{diag2}
\end{eqnarray}
where 
$\psi_{\gamma}$ is a pseudostate wavefunction of pair $\gamma$  
and $\varepsilon_{\gamma}$ is the corresponding pseudostate energy.
Then we can write
\begin{eqnarray}
\Phi &\approx& \sum_{\alpha=1}^{N_{\mathrm{H}}} F_{\alpha}({\bm \rho}_{\alpha})
\psi_{\alpha}({\bm r}_{\alpha}) + \sum_{\beta=1}^{N_{\mathrm{Ps}}}
F_{\beta}({\bm \rho}_{\beta}) 
\psi_{\beta}({\bm r}_{\beta}) 
\nonumber \\ & 
\equiv & \sum_{\gamma=1}^{N_{\mathrm{H}}+N_{\mathrm{Ps}}} F_{\gamma}({\bm \rho}_{\gamma})
\psi_{\gamma}({\bm r}_{\gamma}),
\label{Phi}
\end{eqnarray}
where 
$F_{\gamma}({\bm \rho}_{\gamma})$ is an unknown weight
function, ${\bm r}_{\gamma}$ is the relative position of the particles in
pair ${\gamma}$, ${\bm \rho}_{\gamma}$ is the position of particle ${\gamma}$
relative to the centre of mass (c.m.)  of pair ${\gamma}$ (${\gamma}=\alpha,\beta$), see \Fref{Fig1Jacobi}.  For convenience, here we use the same notation not only to
denote a pair and a corresponding channel, but also a quantum state in
this pair and the channel.  So, the
indices of functions $F_{\gamma}$ and $\psi_{\gamma}$, additionally refer to a
full set of quantum numbers of a state in the channel. In the case of
vectors ${\bm \rho}_{\gamma}$ and ${\bm r}_{\gamma}$, the indices still 
refer only to a channel and a pair in the channel, respectively.

In principle, at this formal stage one could keep the continuum part
only for one of the pairs in order not to double up the treatment of
the three-body breakup channel. However, a symmetric
expansion of the type \eref{Phi}, with the continuum states on both
centres, was found to give fastest convergence in 
calculations with a manageable number of
states~\cite{KB00,KB00jpbl,KB01npa}. 

Now we use the Bubnov-Galerkin  principle~\cite{M64} to find the coefficients $F_{\gamma}({\bm \rho}_{\gamma})$ so that the expansion (\ref{Phi}) satisfies Eq.~(\ref{SEH}) the best possible way. Accordingly, we substitute the expansion (\ref{Phi}) into Eq.~(\ref{SEH})
and require the result to be orthogonal to all
($\gamma'=1,\dots,N_{\mathrm{H}}+N_{\mathrm{Ps}}$) basis states, i.e.
\begin{eqnarray}\label{LS1}
\langle \psi_{\gamma'} | H-E | \sum_{\gamma=1}^{N_{\mathrm{H}}+N_{\mathrm{Ps}}}
F_{\gamma}\psi_{\gamma}\rangle_{\bm{ \rho}_{\gamma'}}  &=& 0.
\end{eqnarray}
In this equation subscript  $\bm{ \rho}_{\gamma'}$ indicates
integration over all
variables except $\bm{ \rho}_{\gamma'}$.
Now taking into account conditions (\ref{diag1}) and (\ref{diag2})
we can write Eq.~(\ref{LS1})
in the following form:
\begin{align}
\label{LS2}
\left(E-\varepsilon_{\gamma'}-\frac{q_{\gamma'}^2}{2 M_{\gamma'}}\right)F_{\gamma'}(\bm{\rho_{\gamma'}})=&
\sum_{\gamma=1}^{N_{\mathrm{H}}+N_{\mathrm{Ps}}}
\left\langle\psi_{\gamma'}
|U_{\gamma' \gamma}|\psi_{\gamma}
\right\rangle 
\nonumber \\ & \times 
F_{\gamma}(\bm{\rho_{\gamma}}).
\end{align}
The potential operators $U_{\gamma' \gamma}$ are given by
\begin{align}
\label{Voper}
U_{\alpha \alpha} &= \overline{V}_{\alpha}, \quad
U_{\beta \beta} = \overline{V}_{\beta}
, \nonumber\\
U_{\alpha \beta}&=U_{\beta \alpha}=H - E.
\end{align}
The condition imposed above in \Eref{LS2} is a system of coupled
equations for unknown expansion coefficients $F_{\gamma}(\bm{\rho_{\gamma}})$. These
functions carry information on the scattering amplitudes.
We transform these integro-differential equations for the weight
functions to a set of coupled Lippmann-Schwinger integral equations for transition amplitudes $T_{\gamma' \gamma}$.

To this end we define the Green's function in channel $\gamma$ 
\begin{eqnarray}
G_{\gamma}(q^2)=\left(E+i0-\frac{q^2}{2M_{\gamma}}-\varepsilon_{\gamma}\right)^{-1},
\label{Green}
\end{eqnarray}
to describe the relative motion of free particle $\gamma$ and 
bound pair $\gamma$  with  binding energy $\epsilon_{\gamma}$. We can now write the formal solution
of the differential equation (\ref{LS2}) as
\begin{align}\label{F-eq2}
F_{\gamma'}(\bm{\rho_{\gamma'}}) =& 
 \delta_{\gamma' \gamma} e^{i{\bm q}_{\gamma}{\bm{ \rho}_{\gamma}}} + \sum_{\gamma''=1}^{N_{\mathrm{H}}+N_{\mathrm{Ps}}}
 \int 
\frac{d{\bm q}}{(2\pi)^3}
{{ e^{i{\bm q}{\bm{ \rho}_{\gamma''}}} }}
{G_{\gamma''}(q^2)}
\nonumber\\&\times
\langle {\bm q}|\left\langle\psi_{\gamma''}
|U_{\gamma'' \gamma}|\psi_{\gamma}\right\rangle|{ F}_{\gamma}\rangle 
.\end{align}
The addition of positive $ i0$
defines the integration path around the singularity point at $q=q_\gamma=[{2M_\gamma(E-\varepsilon_{\gamma})}]^{1/2}$, which is real for $E>\varepsilon_\gamma$, and corresponds to the outgoing  wave boundary conditions.
Taking \Eref{F-eq2} to the asymptotic region one can demonstrate \cite{KMSBP03} that 
\begin{eqnarray}
\label{MEL}
T_{\gamma' \gamma}({\bm q}_{\gamma'},{\bm q}_{\gamma}) =
\langle {\bm q}_{\gamma'}|\left\langle\psi_{\gamma'}
|U_{\gamma' \gamma}|\psi_{\gamma}\right\rangle|{ F}_{\gamma}\rangle
\end{eqnarray}
represents the matrix element for transition from channel ${\gamma}$ to channel ${\gamma'}$,
where $\bm{q}_{\gamma}$ and $\bm{q}_{\gamma'}$ are the initial and final momenta
of the incident and scattered particles.  
Then from (\ref{F-eq2}) we get
\begin{align}\label{F-eq2T}
F_{\gamma'}(\bm{\rho_{\gamma'}}) =& 
 \delta_{\gamma' \gamma} e^{i{\bm q}_{\gamma}{\bm{ \rho}_{\gamma}}} + \sum_{\gamma''=1}^{N_{\mathrm{H}}+N_{\mathrm{Ps}}}
 \int 
 \frac{d{\bm q}}{(2\pi)^3}
{{ e^{i{\bm q}{\bm{ \rho}_{\gamma}}} }}
{G_{\gamma''}(q^2)}
\nonumber\\&\times
T_{\gamma'' \gamma}({\bm q},{\bm q}_{\gamma})  
.\end{align}
Now using Eq.~(\ref{F-eq2T}) in Eq.~(\ref{MEL})
 we get  a set of coupled Lippmann-Schwinger-type equations for the transition matrix elements ($\gamma,\gamma' = 1,...,N_{\mathrm{H}}+N_{\mathrm{Ps}}$)
\begin{align}
T_{\gamma' \gamma}(\bm{q}_{\gamma'},\bm{q}_{\gamma})=&V_{\gamma' \gamma}(\bm{q}_{\gamma'},\bm{q}_{\gamma}) 
+\sum_{\gamma''=1}^{N_{\mathrm{H}}+N_{\mathrm{Ps}}} 
\int \frac{d\bm{q}}{(2\pi)^3}
\nonumber\\& \times 
V_{\gamma' \gamma''}(\bm{q}_{\gamma'},\bm{q})
G_{\gamma''}(q^2)T_{\gamma'' \gamma}(\bm{q},\bm{q}_{\gamma}).
\label{LSorig}
\end{align}
The effective potentials are given by
\begin{eqnarray}
V_{\gamma' \gamma}(\bm{ q}_{\gamma'},\bm{ q}_{\gamma})&=
\langle \bm{ q}_{\gamma'} | \langle \psi_{\gamma'} | U_{\gamma' \gamma}
| \psi_{\gamma} \rangle | \bm{ q}_{\gamma} \rangle .
\label{Veff0}
\end{eqnarray}

Thus we have a set of three-dimensional momentum-space integral equations. \Eref{LSorig} 
can be solved via partial wave decomposition. The latter is an ideal
method for collisions involving light projectiles  like
positrons~\cite{KBS07,BKB15} or
positronium~\cite{KLUB13,Ketal15l}. For heavy projectiles a direct
three-dimensional discretization technique has been developed~\cite{KBSS05,KBS06,KABS09}.

After partial wave expansion for each value of the 
total angular momentum $J$ one gets
\begin{align}
{\cal T}_{\gamma'\gamma}^{L'LJ}(q_{\gamma'}, q_{\gamma}) =&
{\cal V}_{\gamma'\gamma}^{L'LJ}(q_{\gamma'},
q_{\gamma}) 
+ \sum_{\gamma''=1}^{N_{\mathrm{H}}+N_{\mathrm{Ps}}} \sum_{L''} 
\int \frac{d q_{}}{(2 \pi^2)} 
\nonumber \\ & \times 
{\cal V}_{\gamma'\gamma''}^{L'L''J}(q_{\gamma'},q_{})
G_{\gamma''}(q^2_{}) 
{\cal T}_{\gamma''\gamma}^{L''LJ}(q_{},q_{\gamma}),  
\label{LSpw}
\end{align}
where $L$, $L'$ and $L''$ are the relative angular momenta of the 
fragments in channels $\gamma$, $\gamma'$ and $\gamma''$,
respectively.  The effective potentials in partial waves 
are given by
\begin{align}
{\cal V}_{\gamma'\gamma}^{L'LJ}(q_{\gamma'}, q_{\gamma}) =&
\sum_{m'mM'M} \int \int d\widehat{\bm q}_{\gamma'} d\widehat{\bm q}_{\gamma}  
Y^*_{L'M'}(\widehat{\bm q}_{\gamma'}) 
\nonumber \\ & \times 
C^{JK}_{L'M'l'm'}
V_{\gamma'\gamma}({\bm q}_{\gamma'},{\bm q}_{\gamma})
C^{JK}_{LMlm}
\nonumber \\ & \times 
Y_{LM}(\widehat{\bm q}_{\gamma}),
\end{align}
where $C_{LMlm}^{JK}$ are the Clebsch-Gordan coefficients of vector
addition, $Y_{LM}(\widehat{\bm q}_{\gamma})$ is the spherical
harmonics of unit vector $\widehat{\bm q}_{\gamma}$. $l$ ($l'$) is the
angular momenta of pair $\gamma$ ($\gamma'$) and $M$, $m$, $K$ are the
projections of $L$, $l$, $J$, respectively, with $K= M+m
= M'+m'$.

We introduce a K-matrix 
according to
\mbox{${\cal T}={\cal K}/(1+i \pi{\cal K})$}~\cite{BS92}. This reduces \Eref{LSpw} to
a set of equations for the K-matrix
amplitudes
\begin{align}
{\cal K}_{\gamma'\gamma}^{L'LJ}(q_{\gamma'}, q_{\gamma}) =&
{\cal V}_{\gamma'\gamma}^{L'LJ}(q_{\gamma'},
q_{\gamma}) 
+ \sum_{\gamma''=1}^{N_{\mathrm{H}}+N_{\mathrm{Ps}}} \sum_{L''} 
{\cal P} \int \frac{d q_{}}{(2 \pi^2)} 
\nonumber \\ & \times
\frac{{\cal V}_{\gamma'\gamma''}^{L'L''J}(q_{\gamma'},q_{})}
{E - q^2_{}/2M_{\gamma''} - \epsilon_{\gamma''}}
{\cal K}_{\gamma''\gamma}^{L''LJ}(q_{},q_{\gamma}),
\label{LSKmat}
\end{align}
where ${\cal P}$ indicates a principal value integral.
These are the major CCC equations to be solved numerically, in the
form of linear equations $Ax=b$ utilising
only real arithmetic~\cite{BS92}. Upon solution the reconstruction of
the T-matrix and obtaining cross sections is a much simpler
computational task.

Details of calculations of the matrix elements are given by
\citet{KB02}. Here we present the final results in order to highlight
the difference in complexity between the matrix elements for direct
and rearrangement transitions. 
The effective potentials  \eref{Veff0} for direct transitions $\alpha
\rightarrow \alpha'$ 
(transitions between states within the same arrangement) read
\begin{align}
V_{\alpha'\alpha}({\bm q}_{\alpha'},{\bm q}_{\alpha}) 
=& \int d {\bm \rho}_{\alpha} d {\bm r}_{\alpha} 
e^{-i{\bm q}_{\alpha'} {\bm \rho}_{\alpha}} \psi^*_{\alpha'}({\bm r}_{\alpha})  
U_{{\alpha}{\alpha}}
\nonumber \\ & \times 
\psi_{\alpha}({\bm r}_{\alpha}) e^{i{\bm q}_{\alpha} {\bm
\rho}_{\alpha}}.  
\end{align}
In partial waves, after some algebra, we have 
\begin{eqnarray}
{\cal V}_{\alpha'\alpha}^{L'LJ}(q_{\alpha'}, q_{\alpha}) &=&
(4 \pi)^{2} 
(-1)^{J+(L'+L+l'+l)/2}
[L'l]
\nonumber \\  && \times 
\sum_{\lambda} \!^{(2)}
C^{L0}_{L'0 \lambda 0}  
C_{l 0\lambda 0}^{l' 0}  
\left\{ \begin{array}{ccc}
L' & \lambda & L \\
l  &    J    & l'
\end{array} \right\}
\nonumber 
\\  
&& \times 
I_{\alpha'\alpha}^{\lambda}(q_{\alpha'},q_{\alpha}), 
\label{Vdirect}
\end{eqnarray}
where $[l]=\sqrt{2 l +1}$, $[lL]$ is a shorthand for $[l][L]$, and
the braces denote the $6j$-symbol. 
In the sum over $\lambda$ only terms preserving the parity of $l'+l$
survive, leading to increments of 2.
The radial integral is defined as
\begin{align}
I_{\alpha'\alpha}^{\lambda}(q_{\alpha'},q_{\alpha}) =&
\int_0^{\infty} d{\rho}_{\alpha} {\rho}_{\alpha}^2 j_{L'}(q_{\alpha'}
{\rho}_{\alpha})  
j_{L}(q_{\alpha} {\rho}_{\alpha}) 
\nonumber 
\\  
& \times 
\int_0^{\infty} d{r}_{\alpha} {r}_{\alpha}^2 
R_{n'l'}({r}_{\alpha})\, 
{\cal U}^{\lambda}_{{\alpha}{\alpha}}({\rho}_{\alpha},{r}_{\alpha})
 R_{nl}({r}_{\alpha}),
\label{Iraddir}
\end{align}
where
$R_{nl}(r)$ are the radial parts of the pseudostate wavefunctions $\psi$ and
\begin{eqnarray}
{\cal U}^{\lambda}_{{\alpha}{\alpha}}({\rho}_{\alpha},{r}_{\alpha}) =
\left\{  \begin{array}{ll}
\frac{\delta_{\lambda 0}}{{\rho}_{\alpha}} -
\frac{{\rho}_{\alpha}^{\lambda}}{r_{\alpha}^{\lambda+1}} & \mbox{ if 
${\rho}_{\alpha} < r_{\alpha}$} \\ 
\frac{\delta_{\lambda 0}}{{\rho}_{\alpha}} -
\frac{r_{\alpha}^{\lambda}}{{\rho}_{\alpha}^{\lambda+1}} & \mbox{
otherwise}   .
\end{array} \right. 
\end{eqnarray}

Matrix elements for $\beta \rightarrow \beta'$ transitions are 
defined as
\begin{align}
V_{\beta'\beta}({\bm q}_{\beta'},{\bm q}_{\beta}) 
=& \int d {\bm \rho}_{\beta} d {\bm r}_{\beta} 
e^{-i{\bm q}_{\beta'} {\bm \rho}_{\beta}} \psi^*_{\beta'}({\bm r}_{\beta})  
U_{{\beta}{\beta}}
\nonumber \\ & \times 
\psi_{\beta}({\bm r}_{\beta}) e^{i{\bm q}_{\beta} {\bm
\rho}_{\beta}}.  
\end{align}
The corresponding partial-wave amplitudes 
${\cal V}_{\beta'\beta}^{L'L}(q_{\beta'}, q_{\beta})$
have  the same form as expression \eref{Vdirect} but with 
${\cal U}^{\lambda}_{{\beta}{\beta}}({\rho}_{\beta},{r}_{\beta})$ in 
\Eref{Iraddir} for $I_{\beta'\beta}^{\lambda}(q_{\beta'},q_{\beta})$
defined as
\begin{align}
{\cal U}^{\lambda}_{{\beta}{\beta}}({\rho}_{\beta},{r}_{\beta}) =
\left(1-(-1)^{\lambda}\right)
\left \{ 
\begin{array}{ll}
\frac{2^{\lambda+1}{\rho}_{\beta}^{\lambda}}{r_{\beta}^{\lambda+1}} &
\mbox{ if 
${\rho}_{\beta} < 
{r_{\beta}}/{2}$} \\
\frac{r_{\beta}^{\lambda}}{2^{\lambda} {\rho}_{\beta}^{\lambda+1}} &
\mbox{ otherwise} .
\end{array} \right. 
\end{align}

In the case of rearrangement $\alpha \rightarrow \beta$
transitions \Eref{Veff0} for the effective potentials reads as
\begin{align}
V_{\beta\alpha}({\bm q}_{\beta},{\bm q}_{\alpha}) 
=& \int d {\bm \rho}_{\beta} d {\bm r}_{\beta} 
e^{-i{\bm q}_{\beta} {\bm \rho}_{\beta}} \psi^*_{\beta}({\bm
r}_{\beta})   
(H - E)
\nonumber \\ & \times 
\psi_{\alpha}({\bm r}_{\alpha}) e^{i{\bm q}_{\alpha} 
{\bm \rho}_{\alpha}}.  
\end{align}
Calculation of the matrix elements for the rearrangement transitions
is significantly more complicated. Using the Gaussian representation
for the wavefunctions and the interaction 
potentials \citet{HNB90}  calculated the Ps-formation matrix elements for
transition amplitudes between arbitrary H and Ps
states.   \citet{M93} suggested a straightforward way of calculating the matrix elements without using additional expansions.
Following ~\citet{M93}, \citet{KB02} further reduced the
results for the case of the Laguerre-type pseudostates analytically
calculating integrals for momentum-space pseudostate wavefunctions and
formfactors. The final result for the matrix elements in the 
rearrangement channels is written as
\begin{align}
{\cal V}_{\beta\alpha}^{L'L}(q_{\beta}, q_{\alpha}) =&
[l' l L' L][ l'! l!] 
(-1)^{J+L'}
\sum_{l_1'}
\frac{[l_1' l_2']}{[l_1'! l_2'!]} 
2^{-l_1'-1} 
\nonumber \\ & \times
\sum_{l_1}
\frac{[l_1 l_2]}{[l_1! l_2!]} 
q_{\beta}^{l_1'+l_1} 
\sum_{l_1''}\!^{(2)} 
C_{l_10l_1'0}^{l_1''0}
\nonumber \\ & \times
\sum_{l_2''}\!^{(2)} 
C_{l_20l_2'0}^{l_2''0}
\sum_{\lambda}\!^{(2)} 
[\lambda]^2
C_{L'0\lambda 0}^{l_1''0 }
C_{L0\lambda 0 }^{l_2''0}
\nonumber \\ & \times
\left\{\begin{array}{cccccccc}
l_1 &\mbox{}& l &\mbox{}& J &\mbox{}& l'&\mbox{} \\
& l_2 &\mbox{}& L &\mbox{}& L' &\mbox{}& l_1' \\
l_2'&\mbox{}& l_2'' &\mbox{}& \lambda &\mbox{}& l_1''
\end{array}\right\} 
\nonumber \\ & \times 
I_{\beta\alpha}^{\lambda}(q_{\beta},q_{\alpha}) ,
\end{align}
where $[l!]=\sqrt{(2 l +1)!}$, $l_1'+l_2'=l'$ and $l_1+l_2=l$, 
the braces denote the $12j$-symbol of the first kind~\cite{V88}, 
\begin{align}
I_{\beta\alpha}^{\lambda}(q_{\beta},q_{\alpha}) =&
q_{\alpha}^{l_2'+l_2} 
{\cal F}^{(I)}_{\lambda}(q_{\beta},q_{\alpha}) 
+ \frac{1}{\pi q_{\alpha}} 
\int_0^{\infty} dq q^{l_2'+l_2+1}  
\nonumber
\\ 
& \times
Q_L\left(\frac{q^2+q^2_{\alpha}}{2 q q_{\alpha}}\right) 
{\cal F}^{(II)}_{\lambda}(q_{\beta},q),
\end{align}
and where $Q_L$ is a Legendre function of the second kind. For
Legendre polynomial $P_\lambda$ and $z=\widehat{\bm q}_{\beta} \cdot\widehat{\bm q}_{\alpha}$,
\begin{eqnarray}
{\cal F}^{(I,II)}_{\lambda}(q_{\beta},q_{\alpha}) 
= \int_{-1}^1 dz F^{(I,II)}({\bm q}_{\beta},{\bm q}_{\alpha})
P_{\lambda}(z)  
\end{eqnarray}
with
\begin{align}
F^{(I)}({\bm q}_{\beta},{\bm q}_{\alpha}) =&
\left(\frac{q^2_{\alpha}}{2} + \frac{p^2_{\alpha}}{2} -E \right)
\frac{\widetilde{R}^*_{n'l'}({p}_{\beta})   
\widetilde{R}_{nl}({p}_{\alpha})}{p_{\beta}^{l'} p_{\alpha}^{l}}
\nonumber
\\ 
&
+ \frac{\widetilde{R}^*_{n'l'}({p}_{\beta})
\widetilde{u}_{nl}({p}_{\alpha})}{p_{\beta}^{l'} p_{\alpha}^{l}}     
+ \frac{\widetilde{u}^*_{n'l'}({p}_{\beta})   
\widetilde{R}_{nl}({p}_{\alpha})}{p_{\beta}^{l'} p_{\alpha}^{l}},
\label{FI}
\end{align}
and
\begin{eqnarray}
F^{(II)}({\bm q}_{\beta},{\bm q}) =
\frac{\widetilde{R}^*_{n'l'}({p}'_{\beta})   
\widetilde{R}_{nl}({p}'_{\alpha})}
{{{p}'_{\beta}}^{l'} {{p}'_{\alpha}}^{l}}, 
\label{FII}
\end{eqnarray}
where $\widetilde{R}_{nl}({ p})$ and  $\widetilde{u}_{nl}({p})$ are the pseudostate wavefunctions and pseudostate formfactors in momentum
space, respectively. 
Here ${\bm
p}_{\gamma}$ is the relative momentum of the
particles of pair $\gamma$:
\begin{eqnarray}
{\bm p}_{\beta}={\bm q}_{\beta}/2-{\bm q}_{\alpha} \mbox{  and  } 
{\bm p}_{\alpha}={\bm q}_{\beta}-{\bm q}_{\alpha}.
\end{eqnarray}
In addition, ${\bm p}'_{\alpha}$ and ${\bm
p}'_{\beta}$ are the relative momenta of the particles of
the corresponding pairs immediately before and after the 
rearrangement:
\begin{eqnarray}
{\bm p}'_{\beta}={\bm q}_{\beta}/2-{\bm q} \mbox{  and  } 
{\bm p}'_{\alpha}={\bm q}_{\beta}-{\bm q},
\end{eqnarray}
where ${\bm q}$ is the relative momentum of the fragments in
channel $e$.

As mentioned before, 2 sets of pseudostates are obtained by diagonalising the H and Ps Hamiltonians.
The radial parts of pseudostates are taken as
\begin{eqnarray}
R_{nl}({r}) = \sum_{k=1}^{N} B_{nk}^{l} {\xi}_{kl}(r),
\end{eqnarray}
where pseudostate basis ${\xi}_{kl}(r)$ is constructed from the orthogonal Laguerre functions
\begin{eqnarray}
{\xi}_{kl}(r) = N_{kl} (2 r/a)^{l+1} e^{-r/a} 
L_{k-1}^{2l+2}(2 r/a),
\end{eqnarray}
with
\begin{eqnarray}
N_{kl} = \left[\frac{2 (k-1)!}{a (2l+1+k)!}\right]^{1/2}.
\end{eqnarray}
Here $L_{k-1}^{2l+2}(2 r/a)$ are the associated Laguerre polynomials.
Expansion coefficients $B_{nk}^{l}$ are found by diagonalizing the
two-particle Hamiltonian of the relevant pair.   As to the choice of the exponential fall-off parameter, practice shows that
the highest rate of convergence is achieved when $a_{\alpha}$ and
$a_{\beta}$ are chosen to best reproduce the ground state energy of
hydrogen and positronium, respectively, for a given size of the
bases $N_{\mathrm{H}}$ and $N_{\mathrm{Ps}}$.  The choice of $a_{\alpha}=1$ and
$a_{\beta}=0.5$ fulfils this requirement.  Associated pseudostate wavefunctions and form factors in momentum
space used in \Eref{FI} and \Eref{FII} have been calculated in analytic form.

The set of equations \eref{LSKmat} is solved using standard quadrature rules. The singular kernel is discretised using a
Gauss quadrature. A subtraction technique is used to handle moving singularities. Very recently \citet{Bcpc15,Bcpc16} have proposed a
new approach to solving \Eref{LSKmat}, which handles the Green's
function analytically, see Sec.~\ref{analyticG}.

Cross section from initial state $\alpha$ to final state $\alpha'$ (or $\beta$) is calculated
as a cross section for excitation of state $\alpha'$ ($\beta$). The total cross section is obtained as a
sum of all partial cross sections for each state included in
the close-coupling expansion, or utilizing the unitarity of the close-coupling
formalism by applying the optical theorem. 
We calculate  the total cross section in both ways (which should give the same result) to check that the optical theorem is satisfied. 
 The total ionization cross section
is calculated as a sum of
 the integrated cross
sections for positive energy states (of both atom and Ps).
The total Ps-formation cross section is defined as a sum of 
the cross sections for electron capture into Ps bound states.

\subsection{Helium}

A distinct feature of the CCC formalism is that  the resulting set of coupled equations (\ref{LSKmat}) is essentially  independent of the choice of the target, and is  similar to that given  for hydrogen. Therefore,  the basic formalism of
the two-centre CCC method described above can be applied to positron collisions with helium. However, in this case spins play an important role. There are
two target electrons that can form positronium. The electrons can be in spin-singlet or spin-triplet states. Depending on the spin projections of the electron and the positron that form positronium, the latter can be formed in  para
(p-Ps) or ortho (o-Ps) states . The target spin and Ps spin should couple into the total spin that stays unchanged during the scattering process. 
However, separation of the spatial part of the total scattering wavefunction from the spin part is a non-trivial task.

\begin{figure}[htb]
\includegraphics[width=\columnwidth]{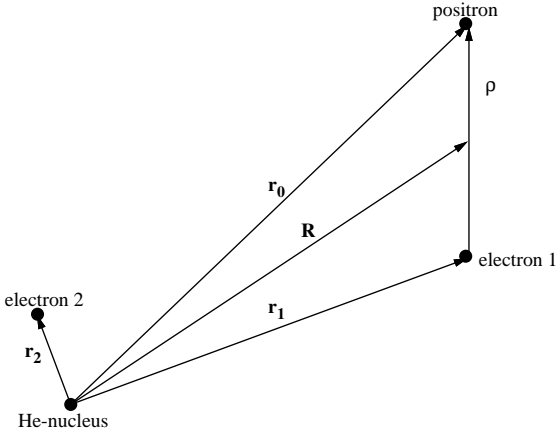}
\caption{Coordinate systems for positron-helium scattering, where we
  assume that positronium is formed with electron 1. In this case the
  generic $R$ and $\rho$ become $R_1$ and $\rho_1$. For the case where
positronium is formed with electron 2 we use $R_2$ and $\rho_2$.}
\label{Coordinate}
\end{figure}

Here it is more convenient to adopt a coordinate system, where ${\bm r_{0}}$, ${\bm r_{1}}$, and ${\bm r_{2}}$ denote the positions of the positron,
electrons 1 and 2, respectively (see Fig.~\ref{Coordinate}), relative to 
the helium nucleus, while
${\bm{R}}=(\bm{r}_{0}+\bm{r}_{1})/2$ is the position of the Ps centre
 relative to the He nucleus and
${\bm{\rho}}=\bm{r}_{0}-\bm{r}_{1}$ is the relative coordinate of the
positron and electron~1. Fig.~\ref{Coordinate} depicts 
one of the two possible systems of coordinates (${\bm r_{0}},
{\bm r_{1}}, {\bm r_{2}}$) and (${\bm{R}}, {\bm{\rho}}, {\bm r_{2}}$). 
There are two sets of Jacobi coordinates corresponding to the two electrons that can form positronium. When necessary we will refer to them
explicitly as (${\bm{R}_1}, {\bm{\rho}_1}, {\bm r_{2}}$) and
(${\bm{R}_2}, {\bm{\rho}_2}, {\bm r_{1}}$). Fig.~\ref{Coordinate}
shows the one 
where Ps is formed by electron 1.

The Schr\"odinger equation for the total scattering wavefunction $\Psi$ of the positron-helium system with the total energy $E$ is written as\begin{eqnarray}\label{SEwiths}
(H-E) \Psi^{sSM}(\bm{ x}_{0},\bm{ x}_{1},\bm{ x}_{2}) = 0 ,
\end{eqnarray}
where $\bm{ x}_{0},\bm{ x}_{1}$ and $\bm{
x}_{2}$ are the full sets of coordinates of the particles including their spin.  
When spin-orbit interactions are neglected it should be possible (at this stage this is merely an assumption; we will see later if this is indeed the case) to separate the radial and spin parts according to
\begin{align}
\label{TW0}
\Psi^{sSM}(\bm{ x}_{0},\bm{ x}_{1},\bm{ x}_{2}) = \Phi^{sS}(\bm{ r}_{0},\bm{
r}_{1},\bm{ r}_{2}) \chi_{s{SM}}(0,(1,2))  , 
\end{align}
where $\Phi^{sS}(\bm{ r}_{0},\bm{
r}_{1},\bm{ r}_{2})$ is the wavefunction that depends on spatial coordinates and $\chi_{s{SM}}(0,(1,2))$ is the spin state of the two electrons (with combined two-particle spin $s$) and the positron with the total spin $S$ and its
projection $M$.  We expand the total wavefunction $\Psi^{sSM}$ in terms
of states of all asymptotic channels and require the result to be antisymmetric against permutation of the two electrons:
\begin{align}
\label{TWFex1}
\Psi^{sSM} \approx& 
\frac{1}{\sqrt{2}}
(1-P_{12})\Big[   \sum^{N_{\rm He}}_{\alpha=1} F^{(s)}_{\alpha}(\bm{ r}_{0})
\psi^{}_{\alpha}(\bm{ r}_{1},\bm{ r}_{2}) 
\nonumber \\ & \times
\chi^{}_{_{s_{} SM}}(0,(1,2))
\nonumber \\&
+ 
 \sum_{s'_{}} \sum^{N_{\rm Ps}}_{\beta=1} 
G^{(s')}_{\beta}(\bm{R}_{1})
\psi_{\beta}(\bm{\rho}_{1})\phi_{1s}^{}(\bm{r}_{2}) 
\nonumber \\ & \times
\chi^{}_{s'_{} SM}((0,1),2) \Big],
\end{align}
where $\psi_{\alpha}$ are the helium wavefunctions and  $\psi_{\beta}$ are the Ps ones. 
Expansion coefficients $F^{(s)}_{\alpha}$ and 
$G^{(s')}_{\beta}$ depend on the spins $s$ and $s'$ of He and Ps,
respectively. The $N_{\rm He}$ and $N_{\rm Ps}$ are the numbers of
the He and Ps pseudostates. The second term represents Ps formation by both electrons and the sum
over $s'$ corresponds to Ps formation in para ($s'_{}=0$)
and ortho ($s'_{}=1$) states. The wavefunction $\phi_{1s}$ describes
the residual ion of He$^+$. The latter is considered to be in the $1s$
state only. The operator $\frac{1}{\sqrt{2}}(1-P_{12})$, where $P_{12}$ is a
permutation operator that interchanges the electrons 1 and 2, insures that the wavefunction is anti-symmetric. 
The spin wavefunctions are written as  \cite{UKFBS10}
\begin{eqnarray}
\label{spa}
\chi^{}_{s_{} SM}(0,(1,2))=\sum_{\mu_{_0},\mu_{}}C^{SM}_{\frac{1}{2}\mu_0\,s_{}\mu_{}} \chi_{\frac{1}{2}\mu_0}(0)
\chi_{s_{}\mu_{}}(1,2)
\end{eqnarray}
for the ${\rm e^+-He}$ channel, and
\begin{eqnarray}
\label{spb}
\chi^{}_{s'_{} SM}((0,1),2)=\sum_{\mu_{_2},\mu'_{}}C^{SM}_{\frac{1}{2}\mu_2\,s'_{}\mu'_{}} \chi_{s'_{}\mu'_{}}(0,1)
\chi_{\frac{1}{2}\mu_2}(2)\cr
\end{eqnarray}
for the ${\rm Ps-He^+}$ channel.

As the spin-orbit interactions are neglected,  the initial spin of the target $s$, the 
total spin $S$ and consequently the total spin function must be conserved. Therefore 
 the spatial part in Eq.~(\ref{TW0}) can be written as 
\begin{eqnarray}\label{spin-sep1}
\Phi^{sS}({\bm r}_0,{\bm r}_1,{\bm r}_2)=\langle\chi^{}_{sSM}(0,(1,2))\,|\, \Psi^{sSM}\rangle 
.\end{eqnarray}
Spin algebra to find the right-hand side of Eq.~(\ref{spin-sep1}) was perform by \citet{UKFBS10pra}.
They showed that the spin wavefunctions satisfy the following
relations
\begin{eqnarray}
\label{spf-a}
\chi^{}_{s_{} SM}(0,(2,1))
&=&  (-1)^{s_{}+1}  \chi^{}_{s_{} SM}(0,(1,2)), 
\end{eqnarray}
\begin{eqnarray}
\label{spf-b}
\chi^{}_{s'_{} SM}((0,1),2)
&=&  \sum_{s''}c_{s'_{}s''S}\chi^{}_{s'' SM}(0,(1,2))
\end{eqnarray}
and 
\begin{align}
\label{spf-c}
\chi^{}_{s'_{} SM}((0,2),1)& =\sum_{s''}(-1)^{s''+1}c_{s'_{}s''S} \chi^{}_{s'' SM}(0,(1,2)), 
\end{align}
where the overlap coefficients are given by the 6j-symbol
\begin{align}
\label{ovlp}
c_{s'_{}s''S} = (-1)^{S-\frac{1}{2}}
[s'
s'']
\left\{ \begin{array}  {cc} {1}/{2}\,\,\,\,\, {1}/{2}\,\,\,\,\, s' \\
S\,\,\,\,\, {1}/{2} \,\,\,\,\, s''_{} \end{array} \right \} 
.\end{align}
Taking into account Eqs. \eref{spf-a}-\eqref{spf-c} and writing the sum over $s'_{}$
explicitly (the Ps spin $s'_{}$ can be 0 or 1) we have
\begin{eqnarray}
\label{TWFex7}
\Phi^{sS} &\approx& \sum^{N_{\rm He}}_{\alpha=1} F^{(s)}_{\alpha}(\bm{ r}_{0})
\psi^{s}_{\alpha}(\bm{ r}_{1},\bm{ r}_{2})
\nonumber \\ && + 
\frac{1}{\sqrt{2}}
\sum^{N_{\rm Ps}}_{\beta=1} \{ \tilde G^{(sS)}_{\beta}(\bm{
R}_{1})\psi_{\beta}(\bm{\rho}_{1})\phi_{1s}^{}(\bm{r}_{2})
\nonumber \\ && + (-1)^{s_{}}  \tilde G^{(sS)}_{\beta}(\bm{ 
R}_{2})\psi_{\beta}(\bm{\rho}_{2})\phi_{1s}^{}(\bm{r}_{1}) 
\} ,
\end{eqnarray}
where 
\begin{eqnarray}\label{asym-hw} 
\psi^{s}_{\alpha}(\bm{ r}_{1},\bm{ r}_{2}) =\frac{1}{\sqrt{2}} 
\{\psi^{}_{\alpha}(\bm{ r}_{1},\bm{ r}_{2})+(-1)^s \psi^{}_{\alpha}(\bm{ r}_{2},\bm{ r}_{1})\}
\end{eqnarray}
is the antisymmetrized helium wavefunction and
\begin{eqnarray}
\label{Gop}
\tilde G^{(sS)}_{\beta}(\bm{R}_{}) = c_{0 s_{}S}  G^{\rm{(0)}}_{\beta}(\bm{R}_{}) + c_{1 s_{}S}G^{\rm{(1)}}_{\beta}(\bm{R}_{}) .
\end{eqnarray}
We emphasize that the superscript  $s$ of $\tilde G_{\beta}$ is the spin of
the target, while that of $G_{\beta}$ is the spin of the formed
positronium.
Note that 
\begin{eqnarray}
G^{(1)}_{\beta}=\sqrt{3} G^{(0)}_{\beta} .
\end{eqnarray}
This relationship is often assumed and used in the literature,
however, its origin remained unclear. It was rigorously derived for the first time by \citet{UKFBS10pra}.

The anti-symmetrised wavefunction $\psi^{s}_{\alpha}(\bm{ r}_{1},\bm{ r}_{2})$ for
helium is built using the
configuration interaction (CI) approach \cite{FB95}. Two types of approximations are used:  a frozen core
(FC), where one
of the electrons is described by the He$^{+}$ 1s orbital and a multi-core (MC), where the core electron is 
described by any number of orbitals as necessary to yield as
accurate He wave
functions as desired.  However, as mentioned earlier the residual ion in the Ps-He$^+$ channels is considered  to be  always in the $1s$-state.  

Eqs.~(\ref{TWFex1}-\ref{TWFex7})  show that 
 the spin part of the total wavefunction is
factorized and, therefore, can be removed from the equations. Consequently, for a given
total spin $S$ and target spin $s$, 
the Schr\"odinger equation (\ref{SEwiths}) transforms to an equation for the spatial part of the total wavefunction $\Phi^{sS}$:
\begin{eqnarray}\label{SE}
(H-E) \Phi^{sS} (\bm{ r}_{0},\bm{ r}_{1},\bm{ r}_{2}) = 0 ,
\end{eqnarray}
where  $\Phi^{sS}$ is given by Eq.~(\ref{TWFex7}).

Direct scattering and Ps-formation matrix elements have been given by \citet{UKFBS10}. Due to the two-centre expansion the system of equations (\ref{LSKmat}) is highly ill-conditioned. The ill-conditioning 
makes it impossible to use arbitrarily high basis sizes, and requires
the matrix elements to be calculated to high precision.

\subsection{Alkali metals}

We model an alkali atom as a system with one active electron
above a frozen Hartree-Fock core \cite{B94}. Accordingly, the positron-alkali
collision is treated as a three-body system of the incoming positron, the active (outer shell) electron
and an inert core ion. 
The interaction between the active electron and the inert core $ V_{\alpha}$
is calculated as a static part of the Hartree-Fock potential $V_{\mathrm{st}}$
supplemented by an exchange potential $V_{\mathrm{ex}}$ between the active and the core
electrons. Thus the interactions in this model system are the electron-ion
$V_{\alpha}$, positron-ion $V_{e}$, and electron-positron $V_{\beta}$
potentials defined as follows
\begin{eqnarray}
  \label{eq:1} 
  V_{\alpha}(r) &=& V_{\mathrm{st}}(r)+V_{\mathrm{ex}}(r)
  +V_{\mathrm{pol}}(r),
 \\
 \label{eq:2}
  V_{e}(r)&=& - V_{\mathrm{st}}(r) + V_{\mathrm{pol}}(r),
 \\
 \label{eq:3}
  V_{\beta}(r)&=&-1/r.
\end{eqnarray}
The
static term is calculated by
\begin{eqnarray}
\label{eq:4} %
  V_{\mathrm{st}}(r) =-\frac{Z}{r}+2 \sum_{\psi_{j}\in C}\int 
    d^{3}r'\frac{|\psi_{j}(\bm{r}')|^{2}}{|\bm{r}-\bm{r}'|},    
\end{eqnarray}
where $Z$ is the charge of the target nucleus and $\psi_j$ are the states of
the ion core $C$ generated by performing the self-consistent-field
Hartree-Fock calculations \cite{B94}. The
summation in Eq.~(\ref{eq:4}) is done for all core states.
The equivalent local-exchange
approximation \cite{FMc73,ABF93,MB81}  is used to take into account the exchange between the active electron and core electrons:
\begin{eqnarray}
  \label{eq:5} %
  V_{\mathrm{ex}}(r,E_{\mathrm{ex}})&=&\frac{1}{2}
  \Big[(E_{\mathrm{ex}}-V_{\mathrm{st}}(r))
\nonumber\\
  &&- \sqrt{(E_{\mathrm{ex}}
   -V_{\mathrm{st}}(r))^{2}+\rho(r)}\Big],
\end{eqnarray}
where 
\begin{eqnarray}
  \label{eq:6} 
  \rho(r)=\sum_{\psi_{j}\in C}\int 
  \mathrm{d}\hat{\bm{r}} |\psi_{j}(\bm{r})|^{2} 
\end{eqnarray}
and $E_{\mathrm{ex}}$ is an adjustment parameter.
The core polarization potential $V_{\mathrm{pol}}$ in Eq.~(\ref{eq:1})
is
\begin{eqnarray}
  \label{eq:7} 
  V_{\mathrm{pol}}(r)=-\frac{\alpha_{d}}{2 r_{0}^{4}} 
  \cdot \frac{1-\exp[-x^{6}]}{x^{4}},
\end{eqnarray}
where $x=r/r_{0}$, $\alpha_{d}$ is the dipole polarizability and
$r_{0}$ are adjustable parameters to fit some physical quantities
(e.g., energies of the valence electron).

\subsection{Magnesium and inert gases}

Mg is modelled as a He-like system with two active electrons
above a frozen Hartree-Fock core \cite{FB97,FB01,SFB11}. The interaction
between an active electron and the frozen Hartree-Fock core
is calculated as a sum of the static part of the Hartree-Fock potential
and an exchange potential between an active and the core
electrons as described in the previous subsection. Details of the two-centre CCC method for positron-Mg collisions are given in \cite{Rav_etal12}.

Wavefunctions for the inert gases of Ne,
Ar, Kr, and Xe are described by a model of six p-electrons above a frozen 
Hartree-Fock core. Discrete and continuum target states are obtained
by allowing one-electron excitations from the p-shell in the following
way. Taking Ne as an example, self-consistent Hartree-Fock calculations are performed for the
Ne$^+$ ion, resulting in the 1s,2s,2p orbitals. The
1s and 2s orbitals are treated as the inert core orbitals, while the
2p Hartree-Fock orbital
is used as the frozen-core orbital to form the target states. A set of Laguerre functions 
is used to diagonalize the quasi-one-electron Hamiltonian of the Ne$^{5+}$ ion,
utilising the nonlocal
Hartree-Fock potential constructed from the inert core
orbitals. The resulting 2p orbital differs substantially from the
Hartree-Fock 2p orbital. A one-electron basis
suitable for the description of a neutral Ne atom is built by
replacing the 2p orbital from diagonalisation with the Hartree-Fock
one, orthogonalized by the Gram-Schmidt procedure.  
The six-electron target states are described via the
configuration-interaction (CI) expansion.
The set of configurations is built by angular momentum
coupling of the wavefunction of 2p$^5$ electrons and the Laguerre-based one-electron
functions. The coefficients of the CI expansion are
obtained by diagonalization of the target Hamiltonian. The target orbital angular momentum $l$,
spin $s$, and parity $\pi$ are conserved quantum numbers and diagonalization
of the target Hamiltonian is performed separately
for each target symmetry \{$l,s,\pi$\}. Full details of the single centre CCC calculations for noble gas
atoms are given in Ref. \cite{FB12_njp}. A two-centre approach to inert
gases has not yet been attempted.

\subsection{Molecular hydrogen}

Positron-H$_2$ scattering can be treated somewhat similar to the helium case.
We consider H$_2$ within the Born-Oppenheimer approximation where the
two protons are considered to be at a fixed internuclear distance
denoted as $d$. Expansion for the total scattering wavefunction (after
separation of the spin part) is similar to \Eref{TWFex7} for
He. However, the wavefunctions for the target  in the first term and
the residual ion in the Ps-formation channel depend on $d$.  The residual ion of H${_2^+}$, with same
internuclear distance $d$, is  considered to be only in its ground state.
We only use a few Ps eigenstates so as to take advantage of their analytical form.
The target states are
obtained by diagonalizing the H$_2$ Hamiltonian
 in a set of antisymmetrized
two-electron configurations, built from Laguerre
one-electron orbitals, for each target symmetry characterized
by the projection of orbital angular momentum,
parity and spin.
To calculate H$_2$ states, 
we use the fixed-nuclei approximation.  Calculations are performed at the ground-state equilibrium
internuclei distance taken to be $d$ = 1.4~a$_0$.
When  $d$ is set to 0 one should obtain the He results. We used this test for both structure and scattering calculations. Details of H$_2$ structure calculations can be found in \cite{ZFB13r,AKFB13l,AKFAB14}.

The derivation of the rearrangement matrix elements are somewhat more
difficult than for He because of their dependence on the nuclear
separation and target orientation. Another difference is that partial
wave expansion is done over the total angular momentum projection $K$. It is convenient to choose the z-axis to be along the $\bm d$ vector (body-frame). Then it is possible to transform the obtained results with this choice of z-axis to any given orientation of the molecule. To facilitate the calculations  only the spherical part of the nuclear potential is considered when 
calculating the rearrangement matrix elements:
\begin{eqnarray}
V_{\rm p}({\bm r}_0,{\bm d})&=&\frac{1}{|{\bm r}_0 - {{\bm d}}/{2}|} + \frac{1}{|{\bm r}_0 + {{\bm d}}/{2}|}
\approx \frac{2}{r_>},
\end{eqnarray}
where $r_{>}={\rm max}\{r_0,d/2\}$.
Then the momentum space representation of the above positron-nucleus potential can be shown to be
\begin{eqnarray}
{\bar V}_{\rm p} (p) = \frac{4\pi^2 {\rm sin}(d 
p)}{d  p^3}.
\end{eqnarray}
With these one further follows the procedure used for positron-He calculations \cite{UKFBS10}. 

For positron collisions with the ground state of H$_2$ only
states with zero total spin are required and so $S=1/2$.
$T$-matrix elements are used for obtaining body-frame scattering
amplitudes. These are then rotated by Euler angles to 
 transform them to lab-frame scattering amplitudes.
Orientationally-independent cross sections are calculated by averaging over all rotations of the molecule \cite{IM96}. 
 An orientationally averaged analytic Born subtraction method
\cite{IM96} is employed for H$_2$ direct transition channels. This helped reduce the number of partial waves
requiring explicit calculation.

\section{Recent applications of the CCC theory to positron scattering}
\subsection{Internal consistency}
\label{intcont}
Fundamentally, in order for a theory to be useful it needs to be
predictive. In the close-coupling approach to electron/positron/proton
scattering on 
relatively simple targets, where the structure is readily obtained, 
there are two computational problems that need to be overcome. The
first, is that for a given set of states used to expand the total
wavefunction the resulting equations need to be solved to an acceptable
numerical precision. The second is to
systematically increase the size of the expansion and demonstrate that the
final results converge to a unique answer that is independent of the
choice of the expansion, so long as it is sufficiently large. Only
once this is achieved can we be in a position to claim that the
results are the true solution of the underlying Schr\"odinger equation
and hence predictive of what should happen in the experiment.

In the case of electron scattering there are only one-centre
expansions because electrons do not form bound states with the
electrons of the target. Electron exchange is handled within the
potential matrix elements all based on the coordinate origin at the
nucleus. Accordingly, establishing convergence in just the one-centre
approach is all that needs to be done, though historically this was a major
challenge~\cite{BS92l,BS92}. Though convergence was shown to be to a
result that disagreed with experiment~\cite{BS92}, subsequent
experiments showed excellent agreement with the CCC
theory~\cite{YCC97,Oetal98}, which lead to reanalysis of the original
data~\cite{WM06b} yielding good agreement with the CCC theory and new
experiments. A 
similar situation occurred in the case of double photoionisation of
helium~\cite{KB98l,Achler01}, which in effect is electron scattering
on the singly charged helium ion~\cite{KB98jpbl,BFKS02}.

For positron and proton scattering the issue of convergence is even
more interesting due to the capacity of the projectile to form a bound
state with a target electron. This leads to a second natural centre in
the problem which also requires treatment to convergence. 
For positron scattering on atomic hydrogen the Ps-formation threshold
is at 6.8~eV, while the ionisation threshold is 13.6~eV. However,
Ps formation is also a form of ionisation of the target except that
the electron is captured to a bound state of the projectile. Any
expansion of the Ps centre using a complete basis will result in
negative- and positive-energy states, with the latter corresponding to
three-body breakup. However, expansion of the atomic centre will also
generate independent positive-energy states corresponding to
three-body breakup. Hence, expansions using a complete
basis on each centre, will yield independent, non-orthogonal
states corresponding to the same physical three-body breakup
process. While this may appear to be a fundamental problem, in
practice it is an interesting strength of the method which allows to
check internal consistency of the results. By this we mean that
the same results must be obtained from a variety of calculations
utilising independent one- and two-centre expansions as detailed below.

We begin by considering two extremes: the first attempts to obtain
convergence using only the atomic centre, while the second attempts
convergence using two complete expansion on both centres. Will either
converge, and if they do, will the convergence be to the same result?

\begin{figure}
\begin{center}
  \includegraphics[width=\columnwidth]{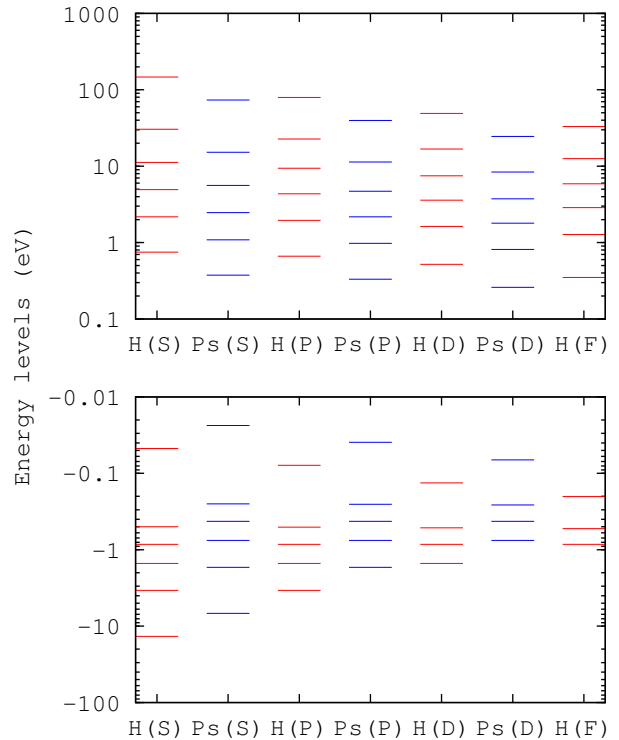}
\end{center}
\caption{Energies $\varepsilon^{\rm (H)}_{nl}$ and $\varepsilon^{\rm
(Ps)}_{nl}$, arising upon diagonalisation of the respective
Hamiltonians, in the CCC(12$_3$,12$_2$) positron-hydrogen 
calculations. Here $12-l$ states were obtained for each $l$ with
$l_{\rm max}=3$ for H states and $l_{\rm max}=2$ for Ps states, see
Kadyrov~\etal~\cite{Ketal15l}. 
\label{energies}      
}
\end{figure}

\begin{figure}
\begin{center}
  \includegraphics[width=\columnwidth]{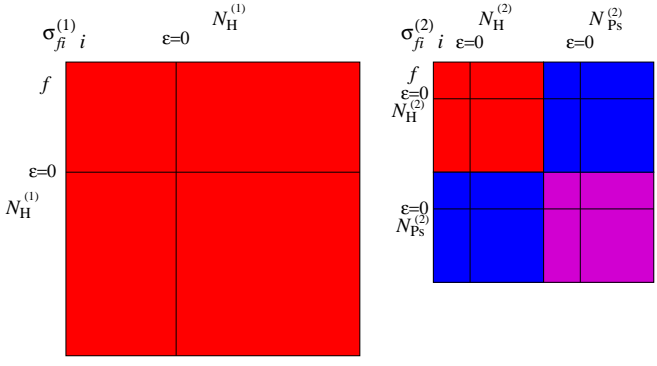}
\end{center}
\caption{Matrix overview
of the cross sections $\sigma_{fi}$ arising from
one-centre (left) and two-centre (right) CCC positron-hydrogen
calculations. $N_{\rm H}^{(1)}$ is the number of H states in the
one-centre calculation which has no explicit Ps states. $N_{\rm
  H}^{(2)}$ is the number of H states in the
two-centre calculation with $N_{\rm Ps}^{(2)}$ explicit Ps states.
First presented by Bray~\etal~\cite{bray2016internal}.
}
\label{picture}    
\end{figure}

In \fref{energies} typical energies arising in two-centre
calculations are given. We see a similar spread of negative- and
positive-energy 
states on both the H and the Ps centres. 
A single centre expansion
based on the atomic centre would not have any Ps states included in
the calculation. 

The results of the two types of calculations may be readily summarised
by \fref{picture}. On the left we have one-centre cross sections
$\sigma_{fi}^{(1)}$, where $i,f=1,\dots,N_{\rm H}^{(1)}$. Taking the
initial state to be the ground state of H ($i=1$) then
$\sigma_{11}^{(1)}$ is the elastic scattering cross section, and
$\sigma_{f1}^{(1)}$ corresponds to excitation whenever
$\varepsilon_{f1}<0$ and ionisation whenever
$\varepsilon_{f1}>0$. The elastic and excitation cross sections need
to converge with increasing $N_{\rm H}^{(1)}$ individually. However,
the ionisation cross sections converge as a sum, yielding the total
ionisation cross section $\sigma^{(1)}_{\rm ion}=\sum_f
\sigma_{f1}^{(1)}$ for $\varepsilon_{f}^{(1)}>0$.

Convergence of the one-centre CCC calculations, where there are no Ps
states, has been studied 
extensively~\cite{BS94,WBFS04jpbl,WBFS04}. Briefly, at energies below the
Ps-formation threshold the important contribution of virtual
Ps formation is adequately treated via the positive-energy atomic
states of large angular momentum $l_{\rm max}\approx 10$. This allows
for convergence of 
elastic scattering cross section to the correct value. At energies
above the ionisation threshold, the positive-energy atomic
pseudostates take into account both breakup and Ps-formation cross
section in a collective way yielding the correct electron-loss and
excitation cross sections. However, in the extended Ore gap region
between the Ps-formation and ionisation thresholds no convergence is
possible due to all positive-energy pseudostates being closed.

Convergence in two-centre calculations is potentially problematic at
all energies due to two independent treatments of the breakup
processes. In practice this manifests itself as an ill-conditioned
system which requires high-precision matrix elements and limits the
size of the calculations i.e. $N_{\rm H}^{(2)}$ and $N_{\rm Ps}^{(2)}$
are typically substantially smaller than $N_{\rm H}^{(1)}$. It is for
this reason that we have drawn the one-center matrix to be
substantially larger in
\fref{picture}. Furthermore, the H-Ps matrix elements 
take at least an order of magnitude longer to calculate
due to the non-separable nature of the radial
integrals~\cite{KB02}. Accordingly, 
even with much smaller number of states (with smaller $l_{\rm max}$)
the two-centre calculations take considerably longer to
complete. Nevertheless convergence is obtained for individual
transitions involving discrete states, explicit Ps formation, and
explicit breakup~\cite{BKB15,bray2016internal}. 
Internal consistency is satisfied if at
energies outside the extended Ore gap, for discrete ($\varepsilon^{\rm
A}_f,
\varepsilon^{\rm A}_i<0$) atomic transitions the two approaches
independently converge such that
\begin{equation}
\sigma_{fi}^{(2)}=\sigma_{fi}^{(1)}. 
\label{disc}
\end{equation}
Furthermore, at energies above the
breakup threshold 
\begin{eqnarray}
\sigma_{\rm eloss}^{(2)}&=&\sigma_{\rm Ps}^{(2)}+\sigma_{\rm brk}^{(2)}\cr
&=&\sigma_{\rm ion}^{(1)},
\label{eloss}
\end{eqnarray}
where for some initial state $i$
\begin{eqnarray}
\sigma_{\rm Ps}^{(2)}&=&\sum_{f:\varepsilon^{\rm
Ps}_f<0}\sigma_{fi}^{(2)},\label{sigps}\\
\sigma_{\rm brk}^{(2)}&=&\sum_{f:\varepsilon^{\rm
Ps}_f>0}\sigma_{fi}^{(2)}+\sum_{f:\varepsilon^{\rm
A}_f>0}\sigma_{fi}^{(2)},\label{sigbrk}\\ 
\sigma_{\rm ion}^{(1)}&=&\sum_{f:\varepsilon^{\rm
A}_f>0}\sigma_{fi}^{(1)}.
\label{sigion}
\end{eqnarray}
In the extended Ore gap only the two-centre calculations are able to
yield convergent results. 

The great strength of the internal-consistency check is that it is
available (outside the extended Ore gap) for every partial wave of the total orbital angular
momentum. Checking that Eq.~\eref{eloss} is satisfied for
every partial wave provides confidence in the overall
results of the two completely independent calculations, which will
typically have very
different convergence properties with increasing $N$ and $l$. Due to the unitarity of the close-coupling
theory agreement for Eq.~\eref{eloss} suggests agreement for other channels, and so
Eq.~\eref{disc} will also hold.

\begin{figure}
\begin{center}
  \includegraphics[width=1.0\columnwidth]{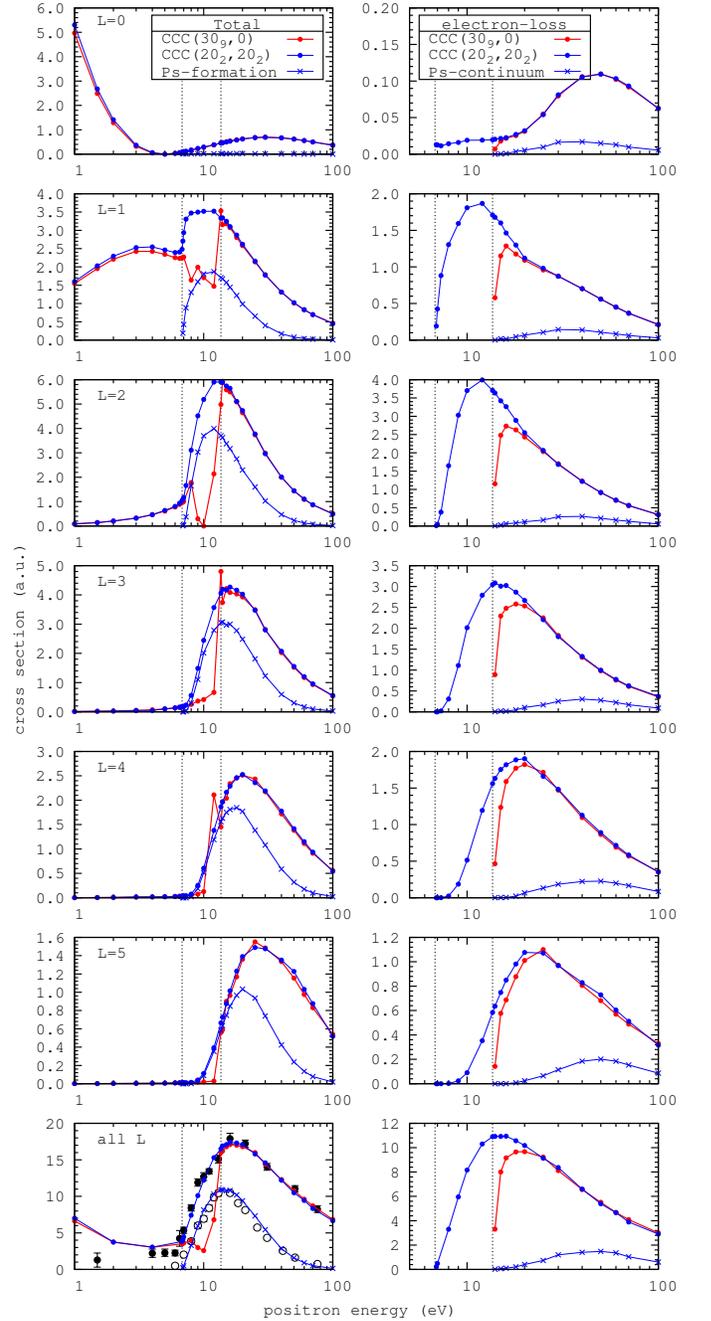}
\end{center}
\caption{Total and electron-loss (Ps formation plus breakup) cross
sections for positron scattering on atomic hydrogen for specified
partial waves $L$ obtained using the one- and
two-centre CCC calculations, see text. The indicated
points corresponding to the energies at
which the calculations were performed are connected with straight
lines to guide the eye. The vertical lines are the Ps-formation and
breakup thresholds, spanning the extended Ore gap. The experimental data in the bottom left panel
are due to \citet{ZLKKS97}. First presented by \citet{bray2016internal}.
\label{p-H} 
}
\end{figure}

In \fref{p-H} we give the example of an internal-consistency check, presented
by \citet{bray2016internal}.
We see that outside the extended Ore gap the two calculations are
generally in
very good agreement. One systematic exception is at just above the ionisation
threshold. Here the breakup cross section is almost zero, but the
Ps-formation cross section is near its maximum. Even with $N_{\rm
H}^{(1)}=30$ the one-centre
calculation does not have enough pseudostates of energy just a little
above zero which would be necessary to reproduce the what should be step-function
behaviour in one-centre CCC calculations. Due to explicit Ps-formation in the
two-centre calculations there are no such problems here or within the
extended Ore gap. Having checked the individual partial waves, and
summing over all to convergence, excellent agreement is found with
experiment. Having performed the internal consistency checks we remain
confident in the theoretical results even if there is potentially a
discrepancy with experiment at the lowest energy measured.

\subsection{Atomic hydrogen}
\label{H-results}
Establishing convergence in the cross sections with a systematically
increasing close-coupling expansion places a severe test on the
scattering formalism. This is as relevant to positron scattering as it
is for electron scattering. Pseudoresonances must disappear with
increased size of the calculations, and uncertainty in the final
results can be established via the convergence study.

One of the earliest successes of the two-centre CCC method for
positron-hydrogen $S$-wave scattering was to show
how the Higgins-Burke pseudoresonance~\cite{HB91} disappeared utilising
a  ($\overline{N},\overline{M}$) basis of only $s$-states on each
centre~\cite{KB00jpbl}. The cross
sections for all reaction channels were shown to  converge to
a few~\% 
with a ($\overline{16},\overline{16}$) basis of
$s$-states. Interestingly, the symmetric treatment of both centres was
particularly efficient in terms of reaching convergence and
eliminating pseudoresonances, with no double-counting problems. 

The  question of convergence in the case of the full
positron-hydrogen scattering problem was investigated by \citet{KB02}. 
Setting $M=N$, states of higher angular momentum were increased
systematically.  The same level of convergence as in the $S$-wave model case was achieved with
the ($\overline{34},\overline{34}$) basis made of ten $s$-, nine $p$-, eight $d$-
and  seven $f$-states for each centre, for scattering on the ground state. The largest calculations performed had a total
of 68 states, 34 each of H and Ps states. The convergence was
checked for the total and other main cross sections corresponding to
transitions to negative-energy states. 
Reasonably smooth cross sections were obtained
for all bases with $l$-convergence being rather rapid.
For the three cases considered $f$-states contribute only marginally.

\begin{figure}[htb]
\begin{center}
\includegraphics[width=70mm,angle=-90]{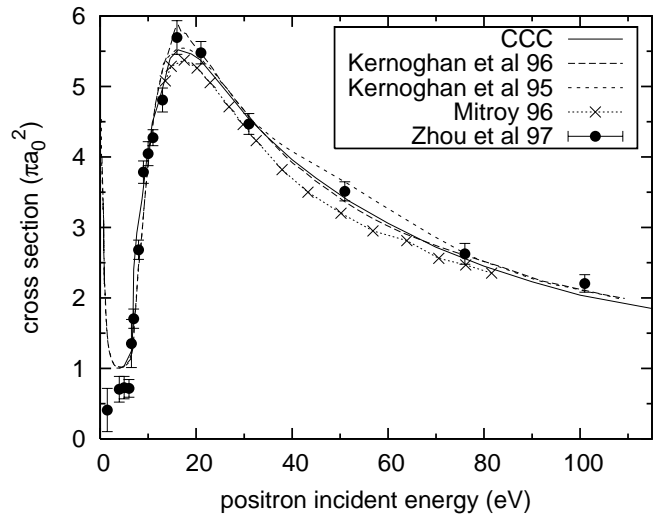} 
\caption{Total cross section for $e^+ +$ H  scattering. The experimental results are due to \citet{ZLKKS97}. The CCC result is from \cite{KB02}. The other theoretical results are due to Kernoghan et al. \cite{KMW95,KRMW96} and \citet{Mitroy96}.}
\label{Fig6}
\end{center}
\end{figure}

\begin{figure}[htb]
\begin{center}
\includegraphics[width=70mm,angle=-90]{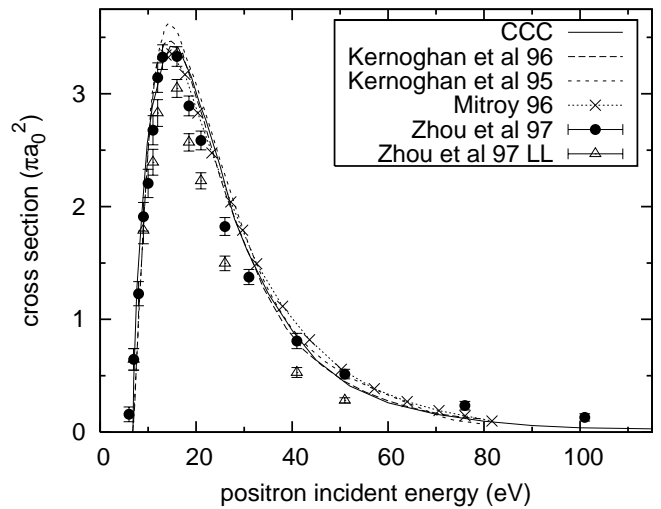} 
\caption{Total Ps-formation cross section for $e^+ +$ H
  scattering. The experimental results are due to \citet{ZLKKS97} (LL
  indicated lower limit). The CCC result is from \cite{KB02}. The other theoretical results are due to Kernoghan et al. \cite{KMW95,KRMW96} and \citet{Mitroy96}.}
\label{Fig7}
\end{center}
\end{figure}

\begin{figure}[htb]
\begin{center}
\includegraphics[width=70mm,angle=-90]{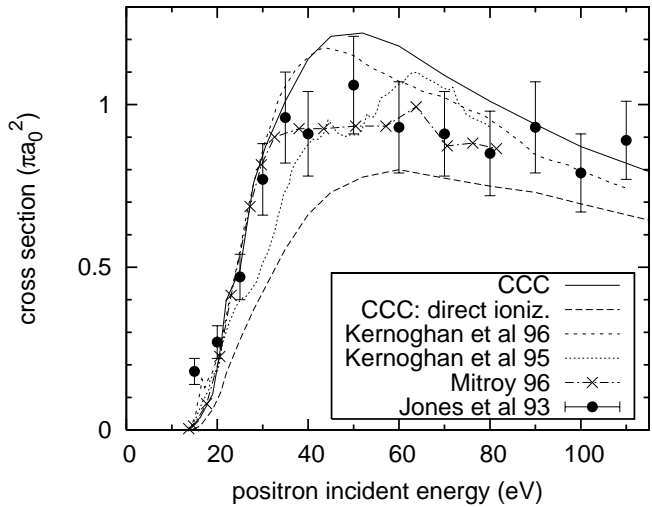} 
\caption{Total breakup cross section for $e^+ +$ H  scattering. The experimental results are due to 
\citet{ucl}. The CCC result is from \cite{KB02}. The other theoretical results are due to Kernoghan et al. \cite{KMW95,KRMW96} and \citet{Mitroy96}.}
\label{Fig8}
\end{center}
\end{figure}

Figs.~\ref{Fig6}-\ref{Fig8} show the CCC results in comparison with
other calculations and experimental data of Detroit~\cite{ZLKKS97} and
London~\cite{ucl} groups. The CCC results agree well with
experiment. So do CC($\overline{30},3$) calculations
of \citet{KRMW96} and CC($\overline{28},3$) calculations of
\citet{Mitroy96}. Note that in these calculations 
the $n^{-3}$ scaling rule was used to estimate the
total Ps formation. Also, an energy-averaging procedure was used to
smooth the CC($\overline{9},\overline{9}$) calculations
of \citet{KMW95}. In the CCC method convergence is established without
such procedures being used.

For the 
breakup cross section the CCC results have two comparable
contributions, one from the excitation of the positive-energy
H pseudostates (shown in \Fref{Fig8} as direct ionization), and the other
from excitation of positive-energy Ps pseudostates. This was also
noted by Kernoghan et al. \cite{KMW95} using the CC($\overline{9},\overline{9}$)
calculations. By contrast, in CC($\overline{N},M$)-type calculations the
contribution to breakup comes only from from direct ionisation. At the maximum
of the cross section the separately converged indirect
contribution to the breakup cross section is approximately a
third of the total. However, the CC($\overline{30},3$) cross section of
\citet{KRMW96} is only a marginally smaller, 
indicating that absence of Ps positive-energy states is absorbed
by the positive-energy H states. 

\begin{figure}[htb]
\centering
\includegraphics[width=\columnwidth]{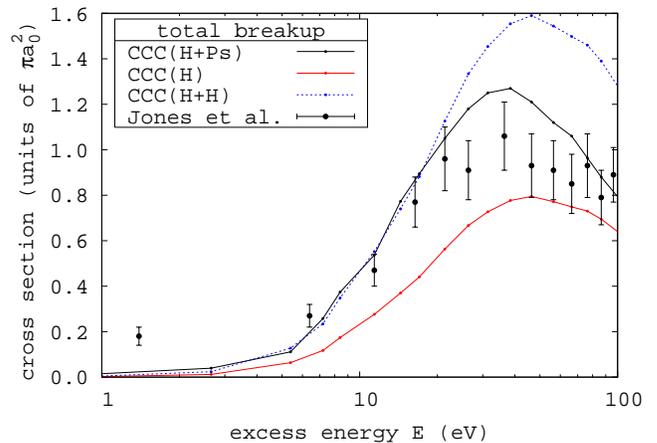} 
\caption{Total $e^+-$H breakup cross section as a function of excess
  energy calculated using the two-center CCC method, first presented
  by \citet{KBS07}.
The argument to the CCC label indicates which center's positive-energy
states were used, see text. The experiment is due to \citet{ucl}.} \label{fig1PRL}
\end{figure}

Fig.~\ref{fig1PRL} shows the CCC results of
\Fref{Fig8}, but against excess (total) energy to
emphasize the lower energies. The full CCC(H+Ps) results with breakup
cross sections coming from both the H and Ps centres are
contrasted with those just from H and twice H (labeled as CCC(H+H)).
We see that below about 20~eV excess energy the
CCC(H+Ps) and CCC(H+H) curves are much the same, indicating that the
Ps and H contributions to breakup converge to each other as the
threshold is approached.

\begin{figure}[htb]
\centering
\includegraphics[width=\columnwidth]{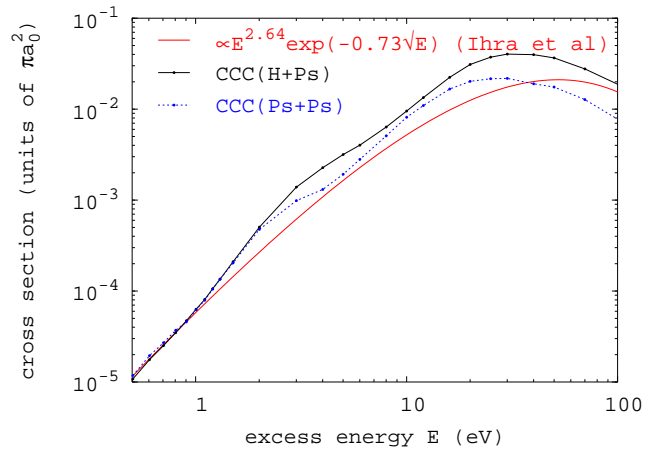} 
\caption{Total $e^+-$H $S$-wave model breakup cross section as a function of excess
  energy calculated using the two-center CCC method, from \citet{KBS07}. As in Fig.~\ref{fig1PRL},
the argument to the CCC label indicates which center's positive-energy
states were used. The Wannier-like threshold law is due to
\citet{Ietal97}}. \label{fig2PRL} 
\end{figure}

Utilising the CCC method \citet{KBS07} 
reported calculations of
positron-hydrogen scattering near the breakup threshold in order to
examine the threshold law.  The results are given in \fref{fig2PRL}.
The Wannier-like threshold law, derived by
\citet{Ietal97}, is in good agreement with the CCC results below
1~eV excess energy. This law was derived for the $L=0$ partial wave,
and \citet{RH94} showed the same energy-dependence holds in
all partial waves. As for the full problem the contributions from both
centres to the breakup cross section converge to each other with
decreasing excess energy, without any over-completeness problems.

\subsection{Helium}
\label{He-results}

Helium in its ground state is the most  frequently used target in experimental studies of positron-atom
scattering. 
First measurements on positron-helium scattering were  carried out by 
 \citet{CCGH72} in 1972.
Since then many other experimental studies have been
conducted~\cite{SKPSJ78,Ketal81,MS94-4349,FKRS86,FDC83,DCBP86,JFKMS95,MAL96}. 
Further developments of positron beams in terms of energy resolution and
beam intensities have recently motivated more experimental
studies~\cite{MSMRL05,MCL09,Karwasz05,KBP10cs,Caradonna_etal09,Caradonna09,SMJCB08,SMJCB08}. 
In general, the results from the experiments agree well with 
each other. A complete theoretical approach from low to high energies
had been lacking until the development of the CCC method for the
problem by \citet{UKFBS10}.

\begin{figure}[htb]
\begin{center}
\vspace{5mm}
\includegraphics[width=\columnwidth, angle=0]{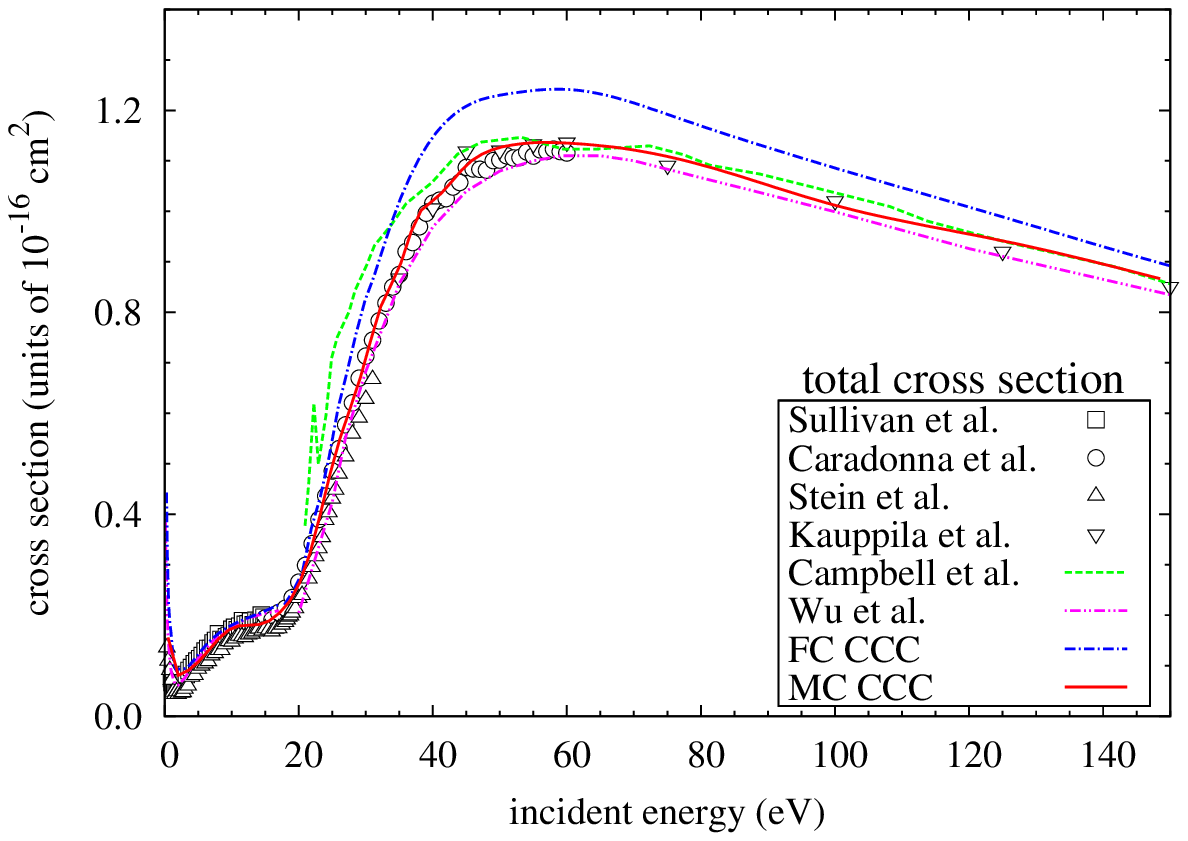}
\caption{Total positron-helium scattering cross section. Experimental data are due to
\citet{SKPSJ78}, \citet{SMJCB08}, \citet{Ketal81}, and
\citet{Caradonna09}. The calculations are due to \citet{CMKW98} and \citet{WBFS04}. The  FC CCC and MC CCC results are from \citet{UKFBS10}. }
\label{TCS-exp}
\end{center}
\end{figure}

A vast amount of experimental data is available for integrated cross sections for positron scattering from the helium ground state. The total CCC-calculated cross section is shown in Fig.~\ref{TCS-exp} in
comparison with experimental data and other calculations. Considerable
discussion on the topic has been presented by \citet{UKFBS10}. It
suffices to say that so long as an accurate ground state is used,
obtained from a multi-core (MC) treatment, agreement with experiment
is outstanding across all energies. We expect the frozen-core (FC)
treatment of helium to result in systematically larger excitation and
ionization cross sections because it understimate the ionization
potential by around 0.84~eV. Generally, a larger ionisation potential leads to a
smaller cross section. In the CCC method we are not free to replace
calculated energies with those from experiment, as this leads to
numerical inconsistency. Consequently, there is no way to avoid the
extra complexity associated with the MC calculations if high accuracy
is required.

\begin{figure}[htb]
\begin{center}
\includegraphics[width=\columnwidth, angle=0]{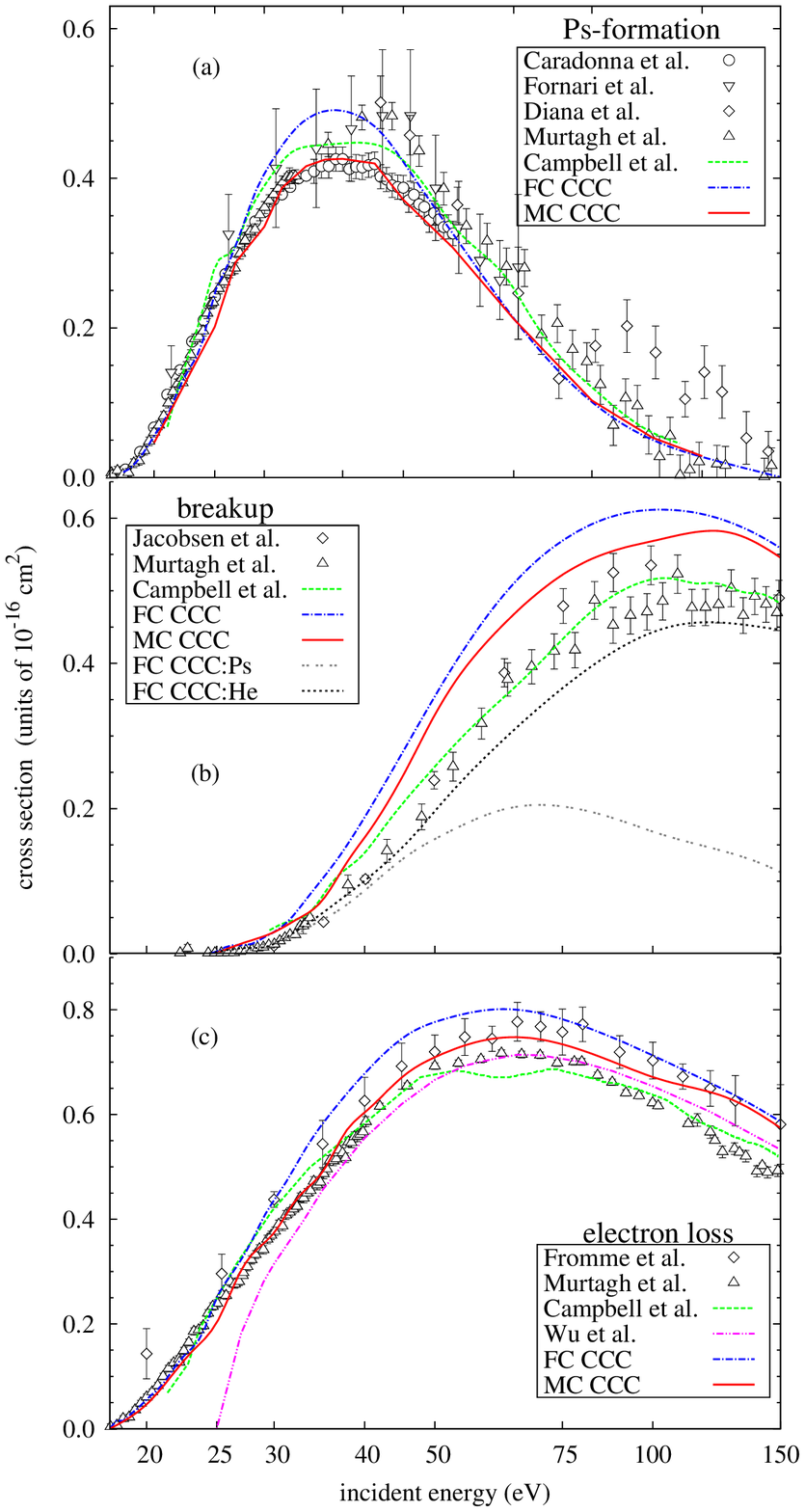}
\caption{Positron-helium (a) Ps-formation, (b) breakup, and (c) electron-loss (sum of
the other two) cross sections. The measurements are due to \citet{Caradonna09},
\citet{FDC83}, \citet{DCBP86}, \citet{MSMRL05}, \citet{FKRS86} and
\citet{KBCP90}. The calculations are due to
\citet{CMKW98}, \citet{WBFS04}. The  FC CCC and MC CCC results are from \cite{UKFBS10}. }
\label{Fig3in1}
\end{center}
\end{figure}

The total Ps-formation, breakup and electron-loss cross sections are
given in Fig.~\ref{Fig3in1}(a), (b) and (c), respectively. 
Beginning with Ps formation, given the minor variation in the
measurements agreement between the various theories and experiment is
satisfactory. However, turning our attention to the breakup cross
section we see that (MC) CCC appears to be substantially higher than
experiment. Yet when these cross sections are summed to form the
electron-loss cross section, the agreement with the experiment of
\citet{FKRS86} which measured this directly,
is good. Given the
complexity of the problem and the experimental uncertainties, the
agreement with experiment is very satisfying.

\begin{figure}[htb]
\begin{center}
\includegraphics[width=\columnwidth, angle=0]{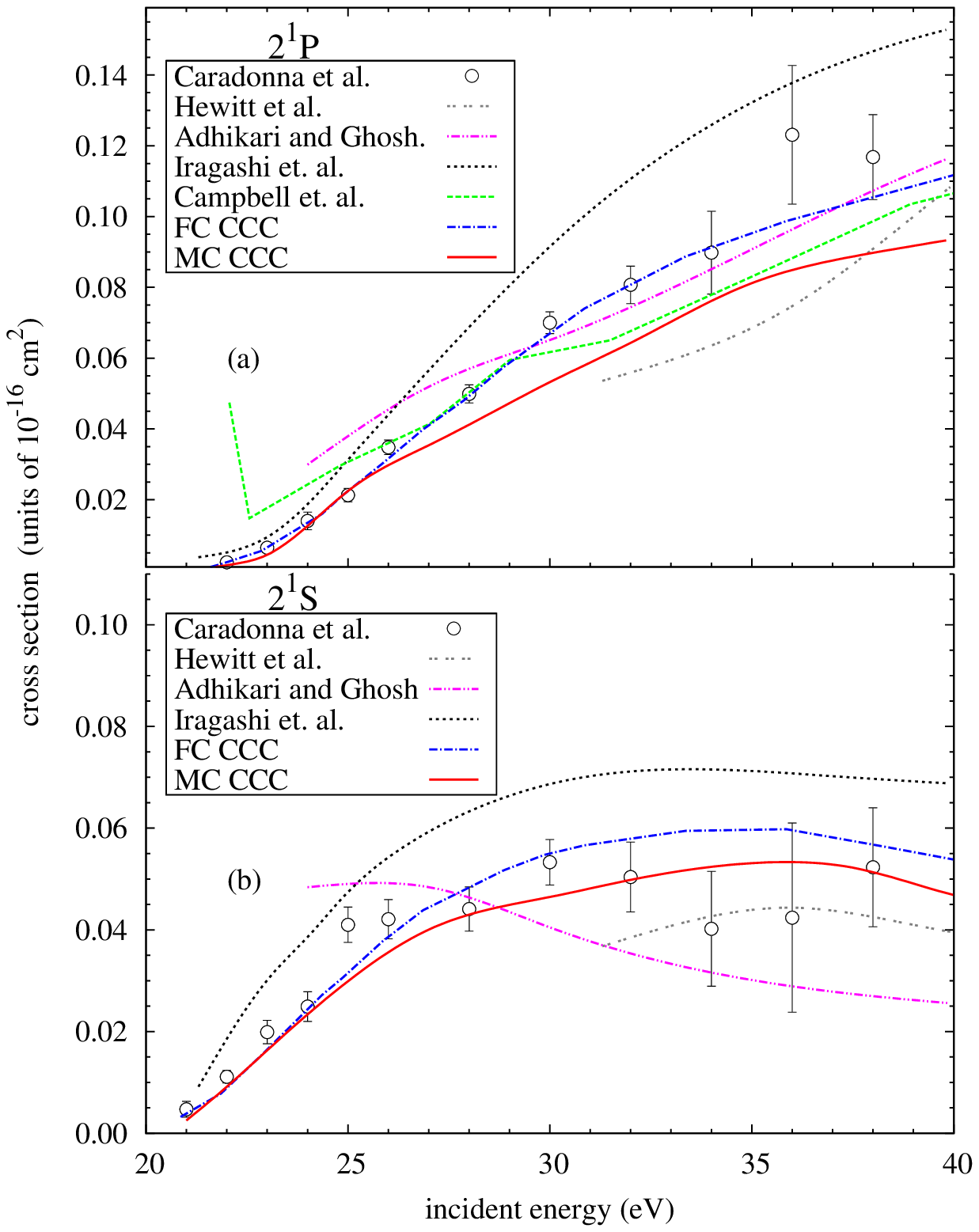}
\caption{Integrated cross sections for (a) He(2$^1$S) and (b)
  He(2$^1$P) excitations by positron impact. Experiment is
due to \citet{Caradonna09}. The  FC CCC and MC CCC results are from \cite{UKFBS10}, other calculations are due to \citet{HNB92},
\citet{AG96}, \citet{ITS96} and  \citet{CMKW98}.}
\label{FigEx}
\end{center}
\end{figure}

The cross sections of 2$^1$S and 2$^1$P excitation of helium 
are presented in Fig.~\ref{FigEx}a and b, respectively.  The MC CCC result for
$2^1$S  is in good agreement with the data of
\citet{Caradonna09} while the $2^1$P result is somewhat lower than
experiment.  The fact that
the FC CCC $2^1$P  results agree better with the experimental data is 
fortuitous. Other available theories show some systematic
difficulties for these relatively small cross sections.

\begin{figure}[htb]
\begin{center}
\includegraphics[width=\columnwidth, angle=0]{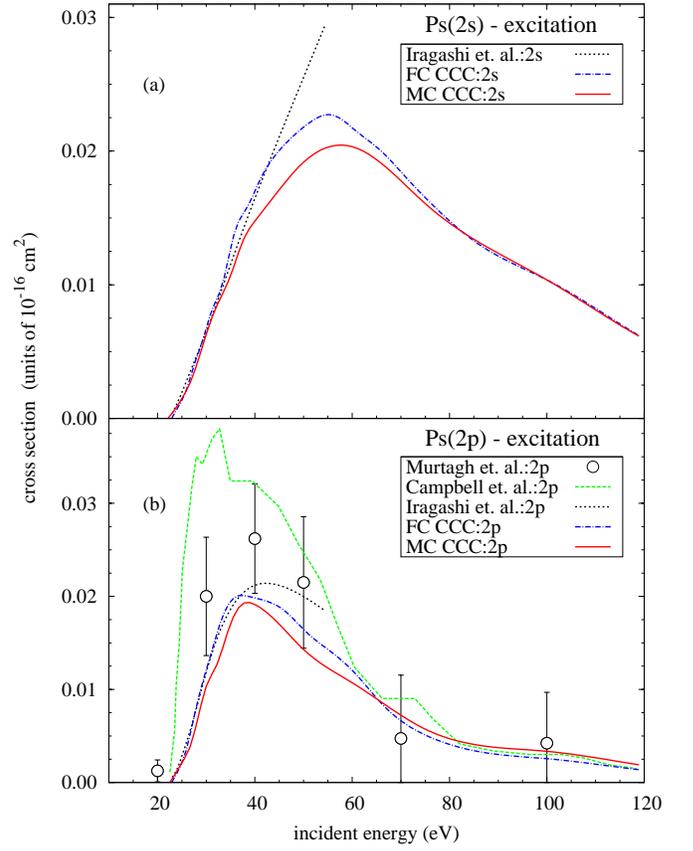}
\caption{Integrated cross sections for  Ps formation in the (a) 2$s$ and (b) 2$p$
states. Experimental data for Ps(2$p$) are due to \citet{MCL09}. The  FC CCC and MC CCC results are from \cite{UKFBS10} and other calculations are due to  \citet{CMKW98} and \citet{ITS96}. }
\label{FigPsEx}
\end{center}
\end{figure}

The cross sections for the rather exotic Ps formation in the 2$s$ and 2$p$ excited states
are presented in Fig.~\ref{FigPsEx}(a) and (b) respectively. 
The cross sections are particularly small. Nevertheless, agreement
with the sole available experiment of  \citet{MCL09} is remarkable.

Thus we have seen that there is good agreement between theory and experiment 
for the integrated cross sections for positron scattering on helium
and hydrogen in their ground states. In both cases the ionisation
thresholds are well above 6.8~eV, and so a one-centre calculation is
applicable at low energies where elastic scattering is the only open
channel. Though experimentally challenging positron scattering on
either H or He metastable excited states results in Ps formation being
an open channel at all energies.  Taking the example of positron
scattering from the $2^3$S metastable state of helium the Ps
threshold is negative (-2.06 eV). This collision system has been
extensively studied by \citet{UKFBS10pra}. As far as we are aware, no experimental 
studies have been conducted for positron scattering on metastable
states of helium. Given the experimental work
on electron scattering from metastable helium \cite{UDTHBB05},  in a
group that also has a positron beam, we are hopeful that in the future there
might be experimental data available for such systems.
As there are still unresolved discrepancies between theory and experiment regarding electron scattering from  
metastable states of He  \cite{LBAL96,Petal98,FB03jpb},  using positrons instead of 
electrons may assist with their resolution.

\subsection{Alkali metals}
\label{alkali-results}
Just like for H and He in metastable states,
for positron collisions with alkali-metal atoms in their ground state, both
elastic and Ps(1s) formation channels are open even at zero positron energy
Accordingly, we require
two-centre expansions even at the lowest incident energies.
\citet{LKBS10}  conducted two-centre calculations with different basis sets to achieve 
results that are independent of the Laguerre exponential
fall-off parameter $\lambda_l$, and convergent with 
the Laguerre basis size $N_l=N_0-l$ for target orbital angular
momentum $l\le l_{\rm max}$. 

\begin{figure}[hbt]
\begin{center}
\includegraphics[width=\columnwidth]{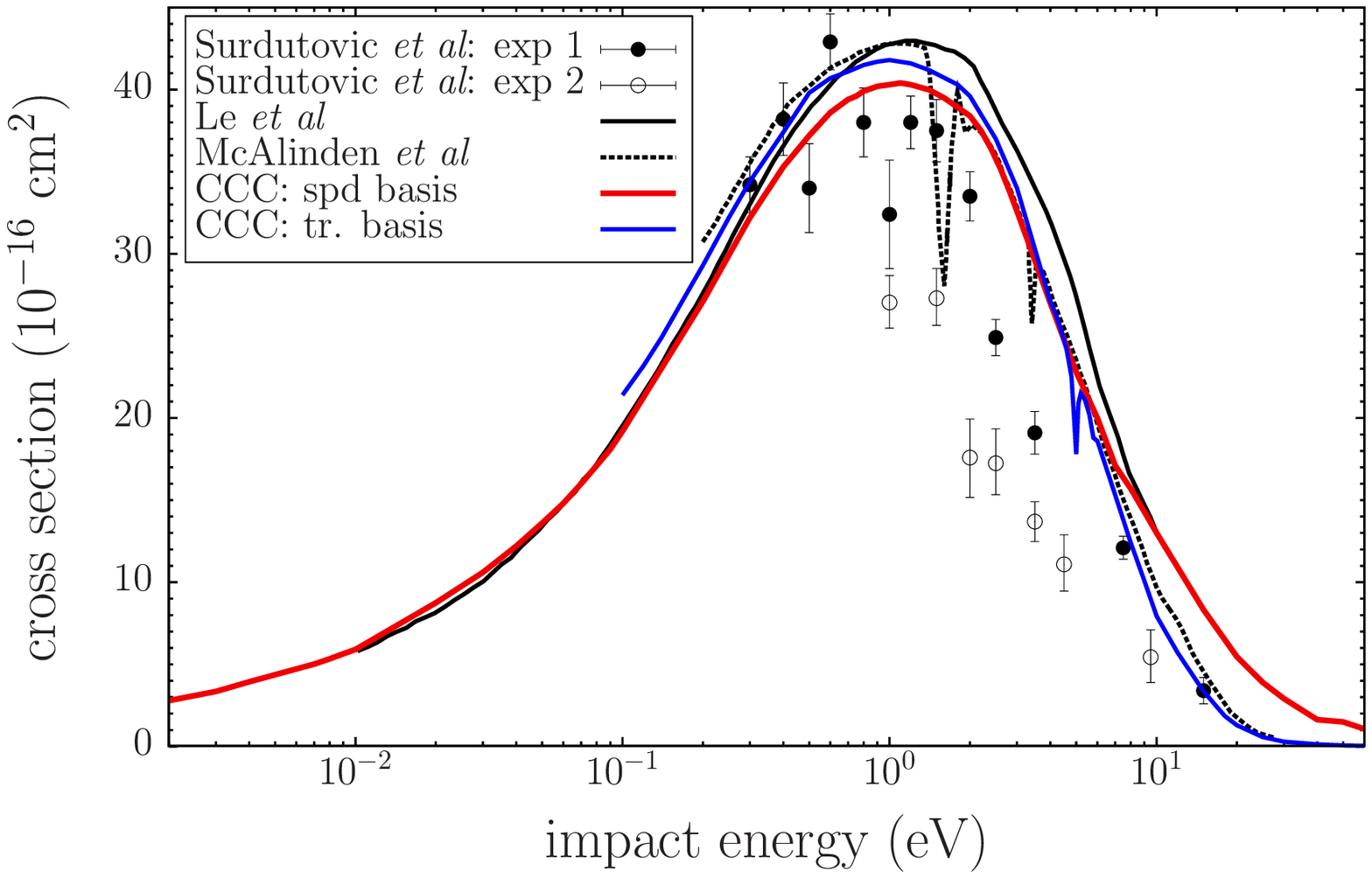}
\end{center}
\caption{Total positronium-formation cross section for e$^+$-Li along
  with the experimental points by \citet{Setal2002} and
  theoretical calculations by
  \citet{McAlindenETAL_1997_p-Li,LeETAL_2005_p-Li-Na}. The truncated basis (CCC
 tr) calculation is an attempt to reproduce the states used by
\citet{McAlindenETAL_1997_p-Li}, see \citet{LKBS10} for details. 
\label{fig:4Li}}
\end{figure}

Figure~\ref{fig:4Li} shows the positronium-formation cross section. 
Agreement between the various calculations is quite good, while
comparison with the 
experimental data of \citet{Setal2002} is rather
mixed. The key feature is that the Ps-formation cross section
diminishes with decreasing energy, supported by all theories, and
consistent with experiment.
Overall, it appears there is no major reason to be concerned.

Unfortunately, changing the target to sodium, substantial discrepancies
between theory and experiment arise, and remain unresolved to
date. One of the motivations for extending the CCC theory to
two-centre calculations of positron-alkali scattering was to address
this problem. \citet{LKBS12}  performed the most extensive study of
this problem that included one- and two-centre calculations. Despite
establishing convergence and consistency of the two approaches no
improvement on previous calculations was found.

\begin{figure}[htb]
  \begin{center}
    \includegraphics[width=\columnwidth]{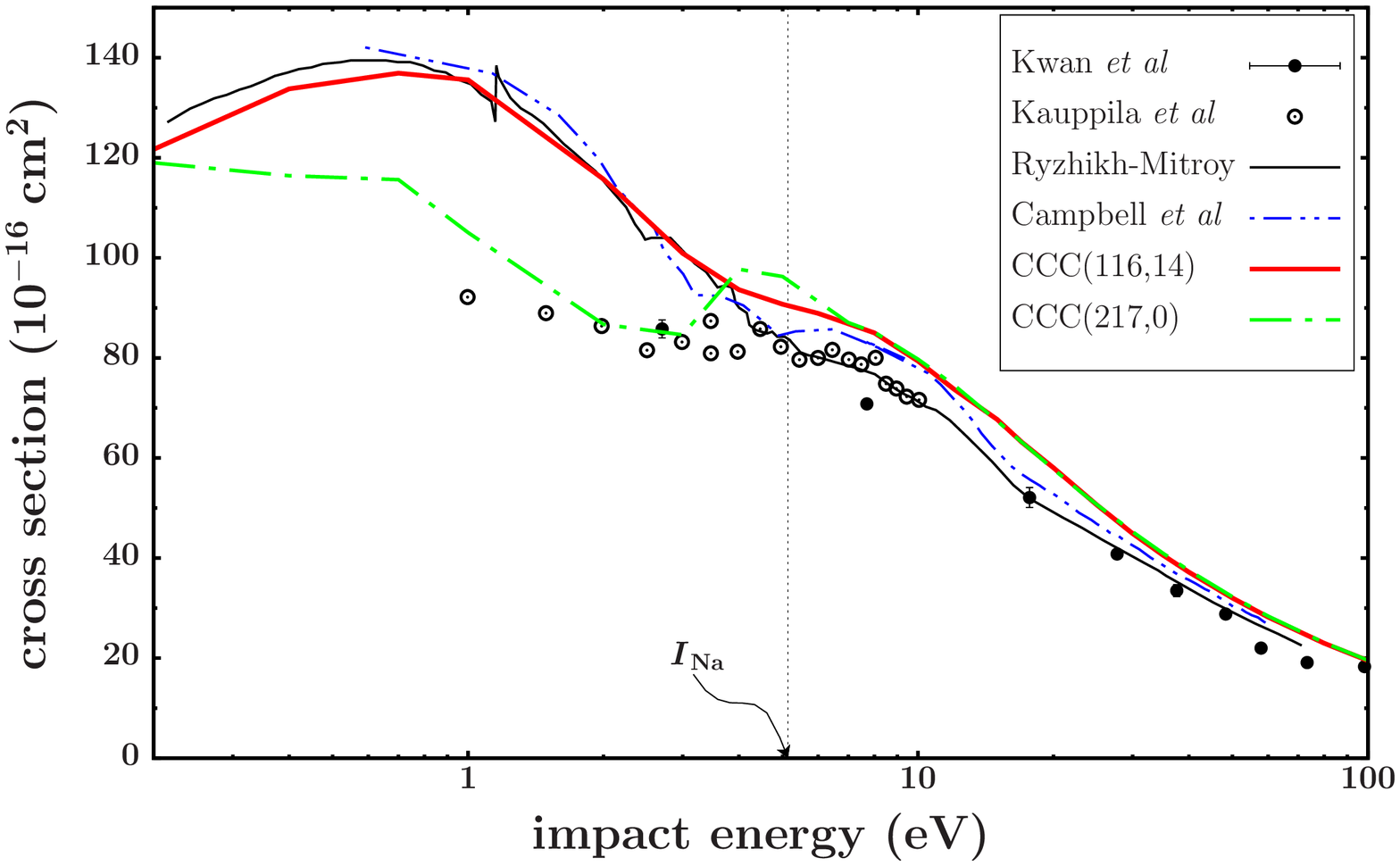}
  \end{center}
  \caption {Total cross section for positron-sodium scattering. The
    two- and one-centre CCC results of \citet{LKBS12} are compared with the calculations of
    \citet{Ryzhikh-Mitroy_1997_JPB_p-Na} and \citet{CMKW98}. The
    experimental points are due to \citet{KwanETAL_1991_TCS-Na-K} and
    \citet{Kauppila-Na-1994}.}
  \label{fig:2Na}
\end{figure}

\begin{figure}[htb]
  \begin{center}
    \includegraphics[width=\columnwidth]{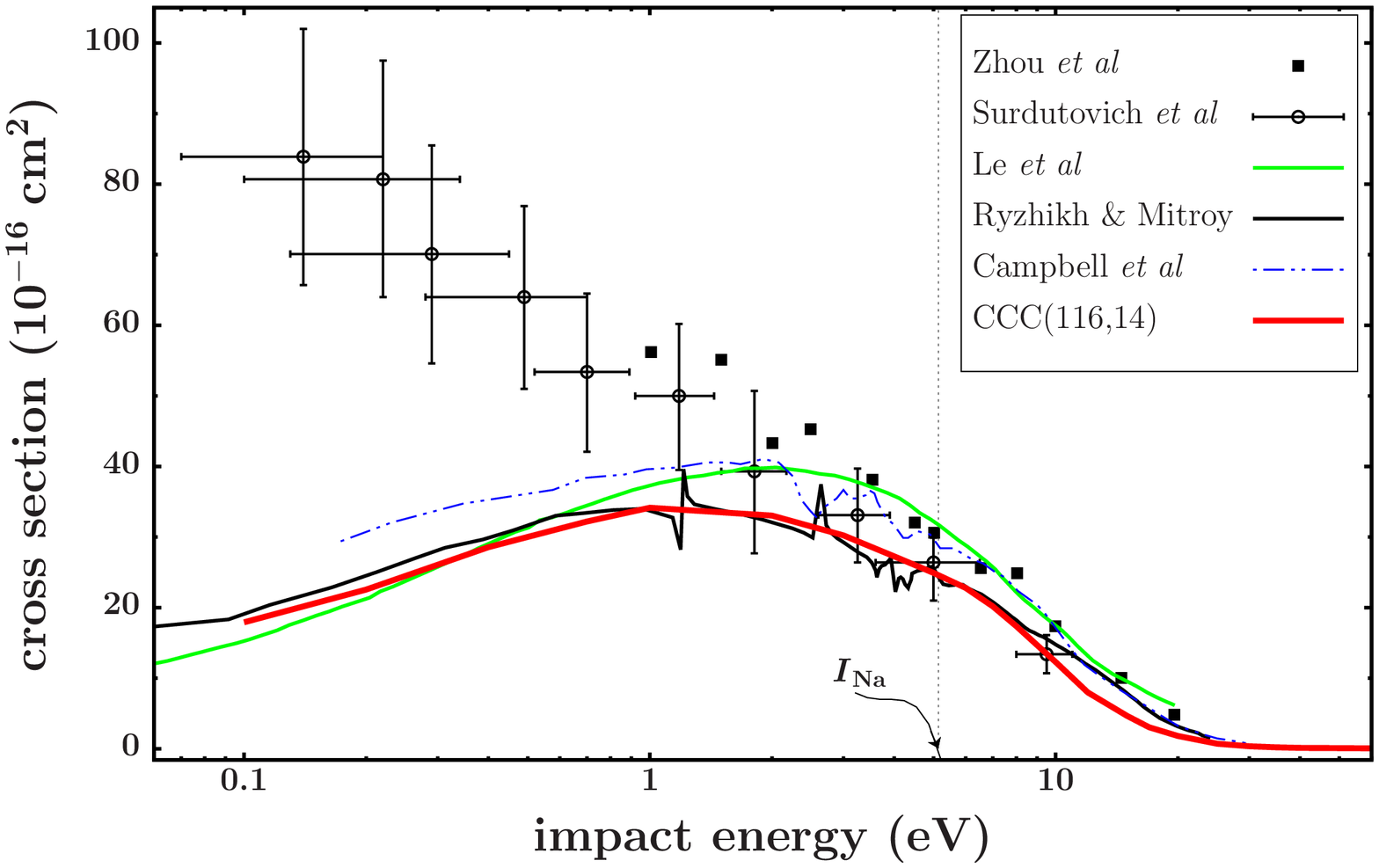}
  \end{center}
  \caption {Total positronium-formation cross section in e$^+$-Na
    scattering. Experiment is due to \citet{Zhou-Ps-form-1994,
      Setal2002}, and the calculations due to
    \citet{Ryzhikh-Mitroy_1997_JPB_p-Na,LeETAL_2005_p-Li-Na} and \citet{LKBS12}. 
  \label{fig:4Na}
}
\end{figure}

We begin by considering the total cross section for positron-sodium
scattering, presented in \fref{fig:2Na}. We see good agreement between
various two-centre calculations, with the one-centre CCC(217,0) calculation
behaving as expected: agreeing with CCC(116,14) only above the
ionisation threshold, and not being valid (or even convergent) below
the ionisation threshold. All of the two-centre calculations are
considerably above the experiment at low energies.

Curiously, the situation is reversed for the total Ps-formation
component of the total cross section, presented in \fref{fig:4Na},
where now all of the theories are considerably below the experiment.
Given that the total cross section at the lowest energies considered
is the sum of elastic and Ps-formation cross sections,
the presented discrepancies with experiment imply that the theoretical
elastic scattering cross sections are overwhelmingly
high~\cite{LKBS12}. Why this 
would be the case remains a mystery.

\begin{figure}
\begin{center}
\includegraphics[width=\columnwidth]{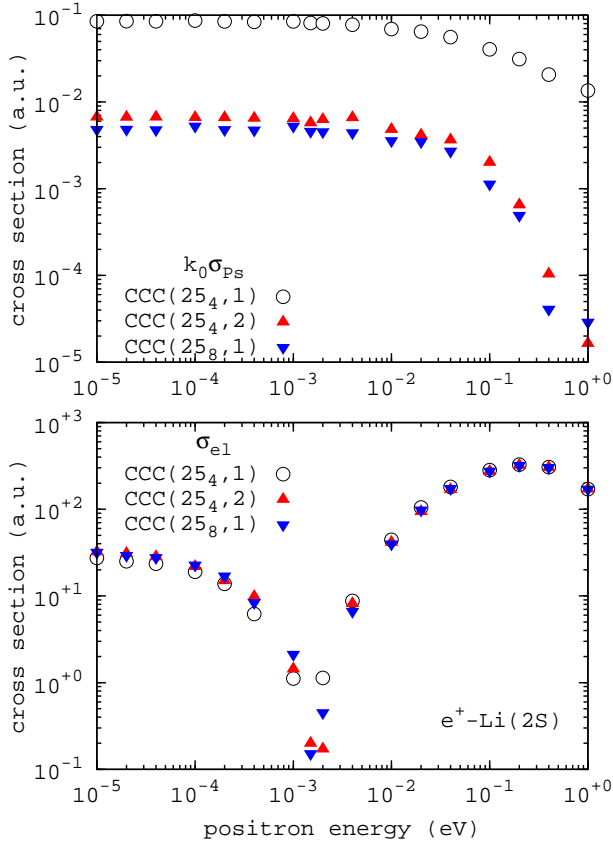}
\end{center}
\caption{Elastic $\sigma_{\rm el}$ (lower panel) and  scaled Ps-formation $k_0 \: \sigma_{\rm Ps}$ (upper panel)  cross sections for positron-lithium scattering, as a function of energy
(13.6$k_0^2$), for the zeroth
partial wave calculated with the indicated CCC($N^{(\rm Li)}_{l_{\rm
max}}, N^{(\rm Ps)}_{l_{\rm max}}$) Laguerre basis parameters, as
presented by \citet{Letal13}. 
\label{Li}
}
\end{figure}

Faced with the problems identified above, \citet{Letal13} considered
threshold behaviour in positron scattering on alkali atoms. 
Following \citet{wigner_behavior_1948} we expect exothermic reactions
such as Ps formation in positron-alkali scattering to result in a cross
section that tends to infinity as $1/k_0$
as the positron energy ($k_0^2$) goes to zero. This is not yet evident in
\fref{fig:4Li} or \fref{fig:4Na} for the considered energies.  Nevertheless, \citet{Letal13} did obtain the
required analytical behaviour, but only in the zeroth partial wave. 
In \fref{Li} the Ps-formation cross section is presented for the zeroth
partial wave multiplied by
$k_0$ so as to demonstrate the expected threshold behaviour. The
convergence study of \citet{Letal13} is also presented, where the
effect of adding the Ps($2s$) state was able to be reproduced by atomic
pseudostates of high orbital angular momentum. \citet{Letal13} found that
higher partial waves $J$ become major contributors to Ps formation at energies above
$10^{-3}$~eV, and have a threshold behaviour as $k_0^{2J-1}$, and so
rise rapidly with increasing energies. It is the contributions of the
higher partial waves that is responsible for the behaviour in the
Ps-formation cross section seen in \fref{fig:4Li}.

\begin{figure}
\begin{center}
\includegraphics[width=\columnwidth]{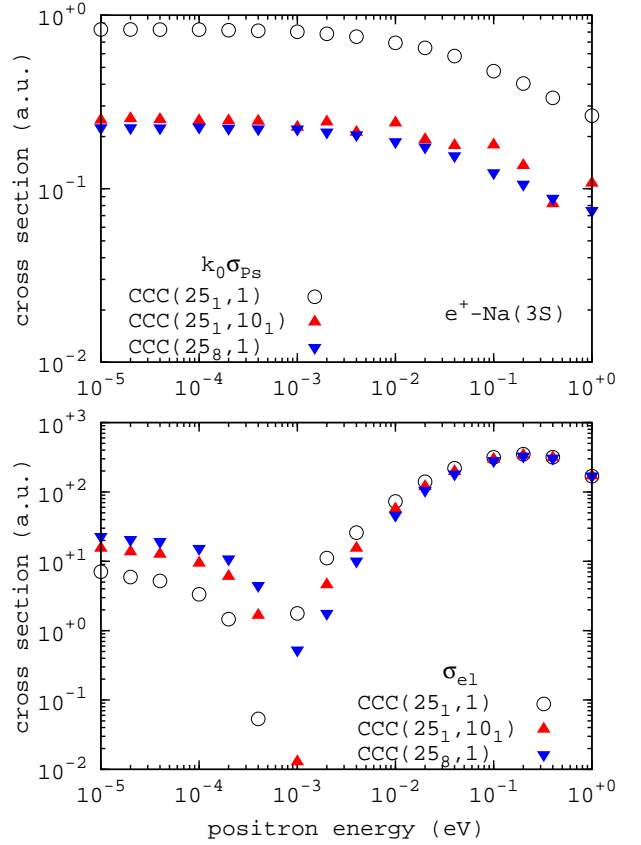}
\end{center}
\caption{The same as Fig.~\ref{Li} but for  positron-sodium scattering.
\label{Na}
}
\end{figure}

We find the same is the case for positron-sodium scattering. In 
\fref{Na} we present  $k_0 \sigma_{\rm Ps}$ and $\sigma_{\rm el}$  for
the zeroth partial wave, demonstrating the expected 
threshold behaviour. Details of the convergence study are discussed by
\citet{Letal13}. It suffices to say that there is a range of
combinations of atomic and Ps states that should yield convergent
results, with such studies being considerably easier at the lower energies
where there are only two open channels. As in the case of lithium the
zeroth partial wave is dominant only below $10^{-3}$~eV, with the
results presented in \fref{fig:4Na} being dominated by the higher partial waves.

\begin{figure}
\begin{center}
\includegraphics[width=\columnwidth]{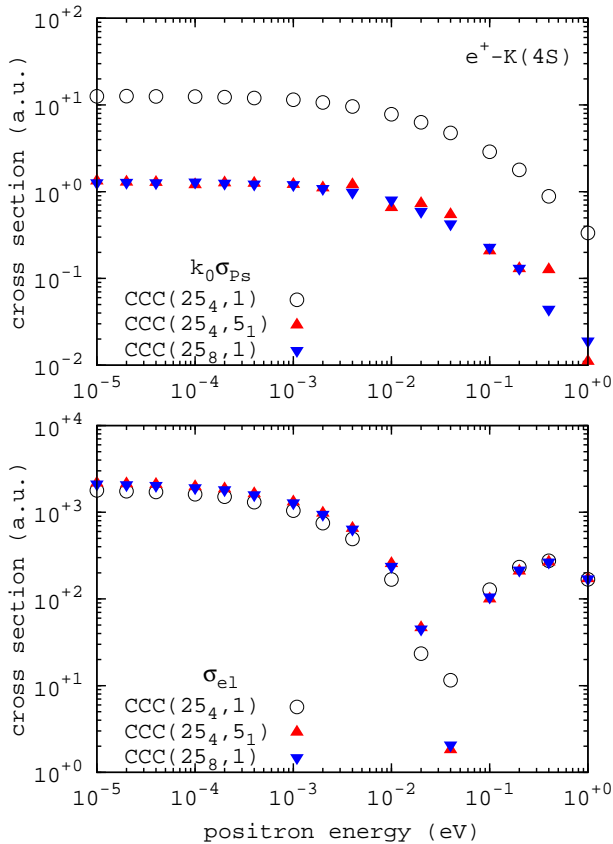}
\end{center}
\caption{The same as Fig.~\ref{Li} but for  positron-potassium scattering.
\label{K}
}
\end{figure}

Lastly, having been unable to explain the discrepancy between
experiment and theory for positron-sodium scattering
\citet{Letal13} 
also considered the positron-potassium scattering system. However,
much the same behaviour as for the lighter alkalis was found, see 
\fref{K}. 
Consequently, the discrepancy between experiment and theory for
low-energy positron scattering on sodium remains unresolved. 

One interesting aspect of the presented elastic cross sections for
positron scattering on the alkalis are the minima. In the case of Li
and Na they are at just above 0.001~eV, whereas for K the minimum is at
around 0.04~eV. Given that in all cases we have just two channels
open, elastic and Ps formation, we have no ready explanation for the
minima, or their positions. The generally smaller Ps-formation cross
section shows no structure in the same energy region, which is
surprising given the substantial minima in the elastic cross section.

\subsection{Magnesium and inert gases}
For experimentalists the transition from say sodium to magnesium for
the purpose of positron scattering is relatively straightforward, not
so in the case of theory. Two valence electrons on top of a
Hartree-Fock core makes for a very complicated projectile-target
combination to treat computationally. However, with an ionization
energy of 7.6~eV the single-centre approach is valid below 0.8~eV
allowing for a test of the two-centre method in this energy
region. This is an important test because in the single centre
approach there are no approximations associated with explicit Ps
formation, and the core is fully treated by the Hartree-Fock approach
rather than an equivalent local core potential approximation.

\begin{figure}[ht]
\includegraphics[width=\columnwidth]{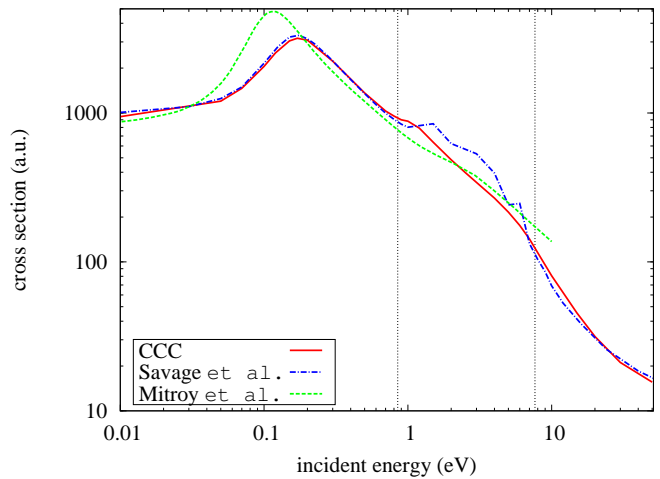}
\caption{e$^+$-Mg elastic-scattering cross section.The first vertical
  line is at the Ps-formation threshold, the second one is at the
  Mg-ionization threshold. The (two-centre) CCC calculations are due
  to \citet{Rav_etal12}, the one-centre due to \citet{SFB11}, and the
  variational ones due to \citet{MB07_prl}. 
\label{Figure3Mg}
}
\end{figure}

One-centre positron-magnesium CCC calculations were presented by
\citet{SFB11}, and two-centre ones by
\citet{Rav_etal12}. The results confirmed the existence of a
low-energy shape-resonance predicted earlier by \citet{Mitroy08}. The
results are presented in \fref{Figure3Mg}. Given that the resonance is
at a very low energy and that a slight energy difference in the target
structure may affect its position the agreement between the theories
is very encouraging. As explained in Sec.~\ref{intcont}, one-centre calculations are unable to
yield convergent results within the extended Ore gap (presently between 0.8~eV
and 7.6~eV). The unphysical
structures displayed within this energy region in the calculations of
\citet{SFB11} depend on the choice of basis with just one example presented.

\begin{figure}[ht]
\includegraphics[width=\columnwidth]{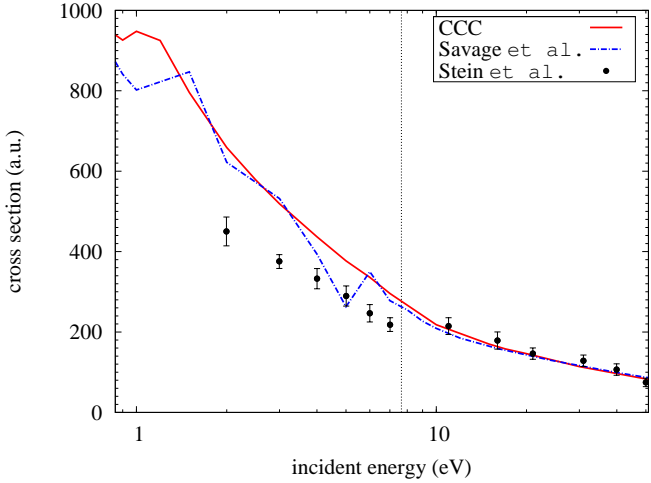}
\caption{e$^+$-Mg total scattering cross section. The vertical line
  indicates the Mg-ionization threshold. Experimental data are due to
  \citet{Stein98}, and the calculations are as for \fref{Figure3Mg}. }
\label{Figure4Mg}
\end{figure}

Unfortunately, the agreement with the experiment of \citet{Stein98}
for the total cross section has the unexpected feature of being good
above the ionisation threshold and poor below, see
\fref{Figure4Mg}. Given that the validity of the
two-centre CCC approach should be energy independent we are unable
to explain the discrepancy. 

\begin{figure}[t]
\includegraphics[width=\columnwidth]{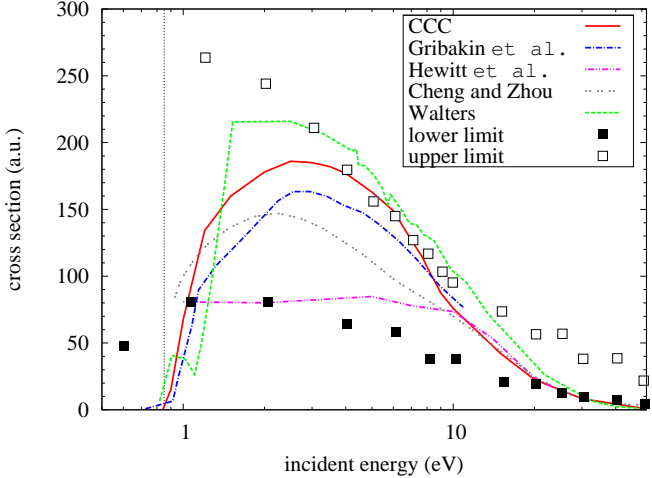}
\caption{e$^+$-Mg Ps-formation cross section. Experimental data for
  upper and lower limits of the Ps-formation cross section are due to
  \citet{SHKKS03}. The vertical line indicates the position of the
  Ps-formation threshold. The calculations are due to
  \citet{Rav_etal12}, \citet{Gribakin96}, \citet{HNBJ96}, Walters (data taken from Ref.~\cite{SHKKS03}) and \citet{CZ06}.
\label{Figure5Mg}
}
\end{figure}

The Ps-formation cross section is presented in
Fig.~\ref{Figure5Mg}, where there are large experimental uncertainties. 
The experimental data of \citet{SHKKS03}
are preliminary estimations for the upper and the lower limits
of the Ps-formation cross section.  The (two-centre) CCC results are
compared with the previous calculations. While there is substantial
variation the theories tend to favour the upper limit estimates.

Positron scattering on inert-gas atoms has been studied using the
single-centre CCC method~\cite{FB12_njp}. As discussed in the introduction
the complexity of adding the second (Ps) centre is immense, and has not
yet been attempted within a convergent close-coupling formalism. The
large ionisation thresholds mean that the 
single centre calculations are valid for elastic scattering on the
substantial energy range
below the Ps-formation threshold, as well as above the
ionisation threshold, but no explicit Ps-formation cross section may
be determined. This is particularly unfortunute in light of the intriguing
cusp-like behaviour observed by \citet{JCMSMMSB10} across the
Ps-formation threshold. Due to the small magnitude of the structures a
highly accurate theoretical treatment is required, but does not yet exist.
The experimental and
theoretical situation for positron-noble gas collision systems
has been recently reviewed extensively by \citet{CZ14}.

\subsection{Molecular hydrogen}
\label{H2-results}
If we allow the internuclear separation of the two protons in H$_2$ to
be fixed then the extra complexity, relative to the helium
atom, is somewhat manageable. 

Various cross sections for $e^+-{\rm H}_2$ scattering have been
recently calculated by \citet{Uetal15} using the two-center CCC
method. This represents the most complex implementation of the
two-centre formalism to date. Issues regarding convergence with
Laguerre-based molecular and Ps states have been
discussed in some detail. Calculations with only
up to three
Ps($1s$), Ps($2s$) and Ps($2p$) states on the Ps centre were presented
above 10~eV. At lower energies the current implementation of the
two-centre formalism fails to pass the internal consistency check
with the single-center calculations of \citet{ZFB13r}. The low
energies are particularly sensitive to the approximations of the
treatment of the (virtual) Ps formation in the field of the highly
structured H$_2^+$ ion.
Here we just present the
key two-centre results in comparison of experiment and theory.

\begin{figure}[htb]
\begin{center}
\includegraphics[width=\columnwidth, angle=0]{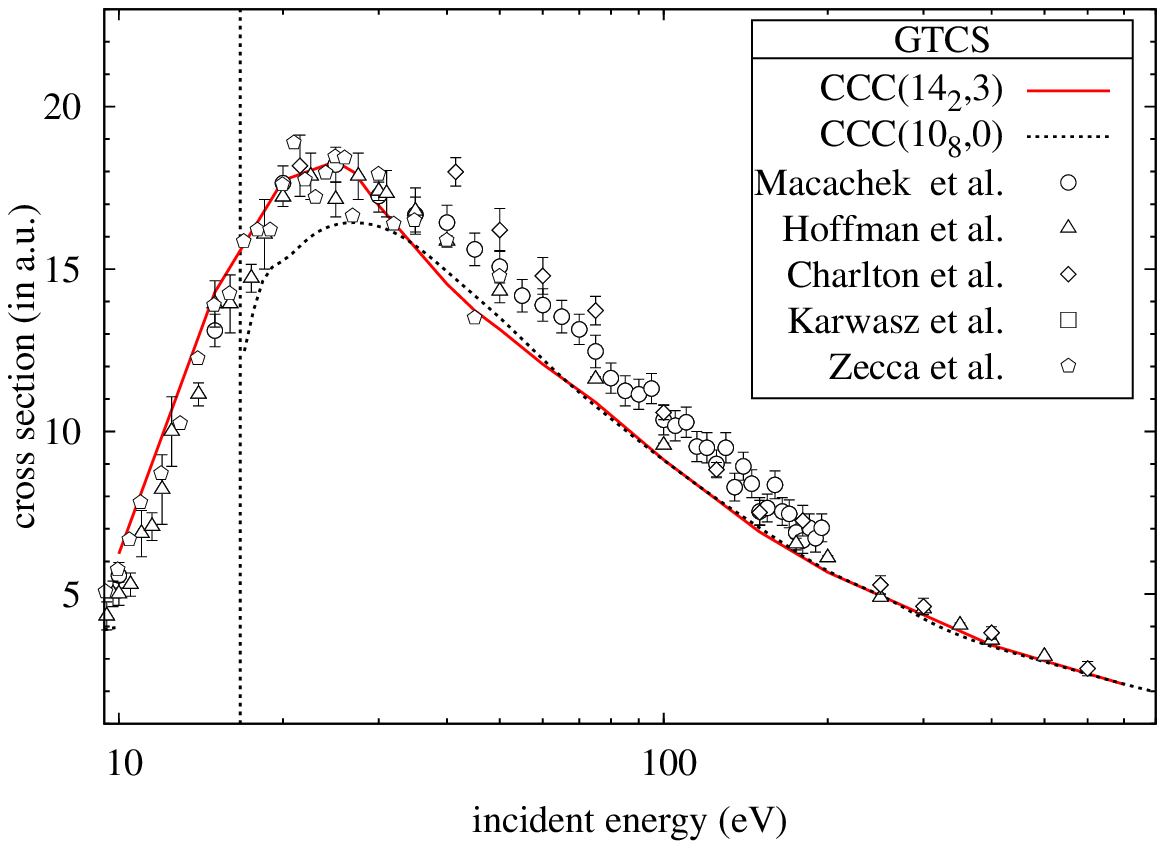}
\caption{The grand total cross section (GTCS) for $e^+-$H$_2$. Experimental data are due to
 \citet{MAMBS2013}, \citet{HDHKPSS82}, \citet{CGHW83}, \citet{KPB06}
 and \citet{ZCSNB09}. The single centre CCC($10_8,0$) results are due to
 \citet{ZFB13r}. The two-centre CCC calculations are due to \citet{Uetal15}.
\label{Figure3H2}
}
\end{center}
\end{figure}

In Fig.~\ref{Figure3H2} the theoretical results are compared  with the
available experimental data for the grand total cross section. Good
agreement between the two- and one-centre calculation above the
ionisation threshold is very satisfying, even if in this energy region
the theory is somewhat below experiment. Good agreement with
experiment of the two-centre CCC results below the ionisation threshold,
dominated by the elastic and Ps-formation cross sections, is
particularly pleasing.

\begin{figure}[th]
\begin{center}
\includegraphics[width=\columnwidth, angle=0]{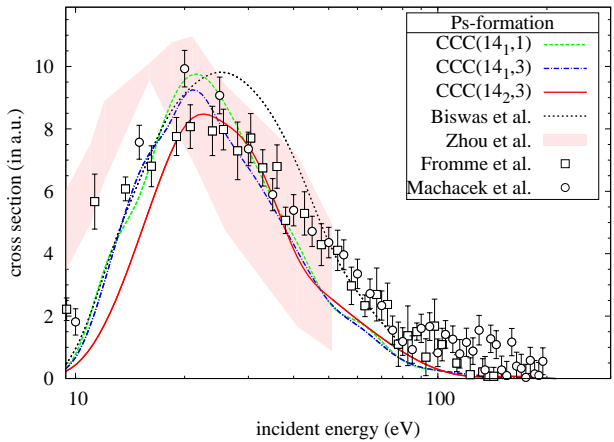}
\caption{The Ps-formation cross section in $e^+-$H$_2$ collisions. Experimental data are due to
\citet{ZLKKS97} (shaded region incorporates upper and lower limits
with their uncertainties), \citet{FKRS88} and \citet{MAMBS2013}. 
Coupled static model calculations are due to \citet{BMG91}. The CCC
calculations are due to \citet{Uetal15}.
\label{Figure4H2}
}
\end{center}
\end{figure}

Fig.~\ref{Figure4H2} shows the
Ps-formation cross section which is a substantial component of the
GTCS, particularly near its maximum around 20~eV.
There is a little variation
between the three CCC calculations particularly at the lower
energies, with the largest CCC calculation being uniformly a little
lower than experiment.

\begin{figure}[th]
\begin{center}
\includegraphics[width=\columnwidth, angle=0]{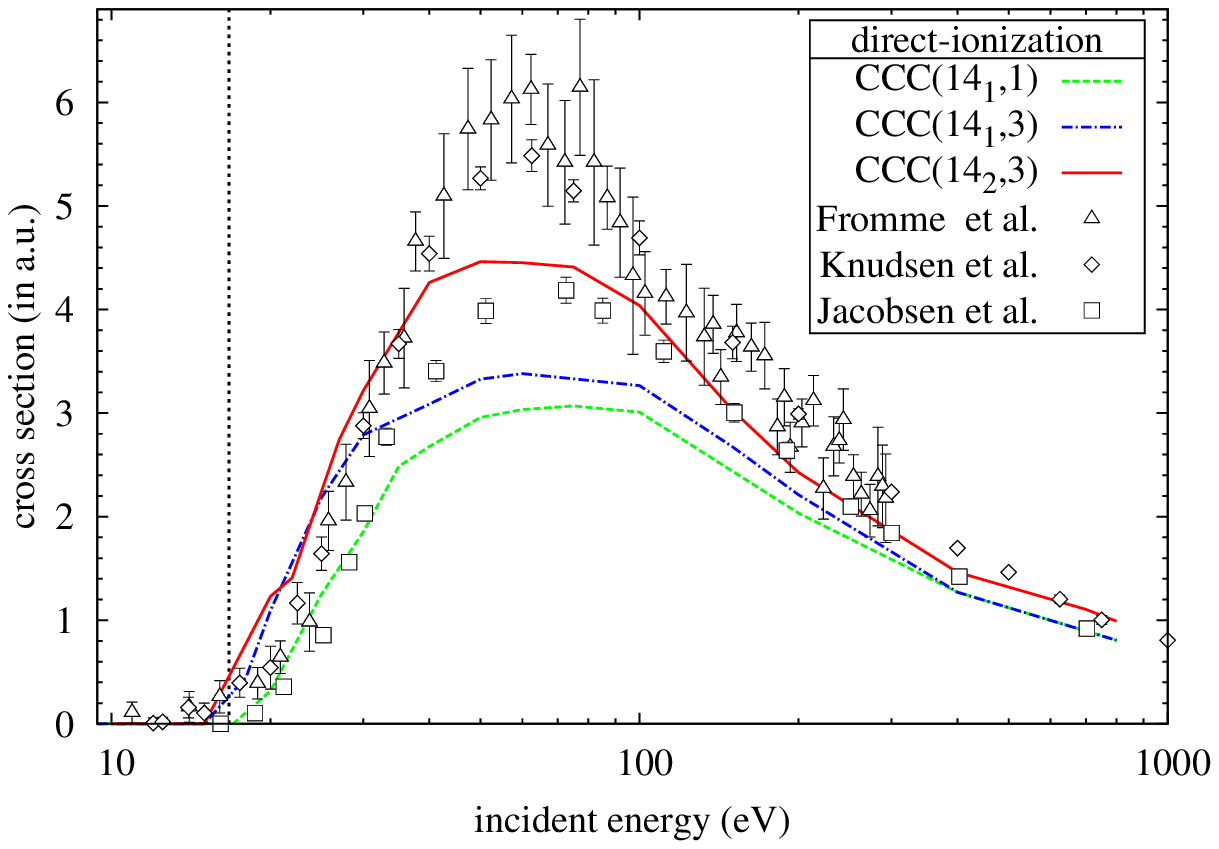}
\caption{The total direct-ionization cross section for $e^+-$H$_2$ collisions. Experimental data are due to
\citet{FKRS88}, \citet{KBCP90} and \citet{JFKM95}. The CCC
calculations are due to \citet{Uetal15}.
\label{Figure5H2}
}
\end{center}
\end{figure}

\begin{figure}[th]
\begin{center}
\includegraphics[width=\columnwidth, angle=0]{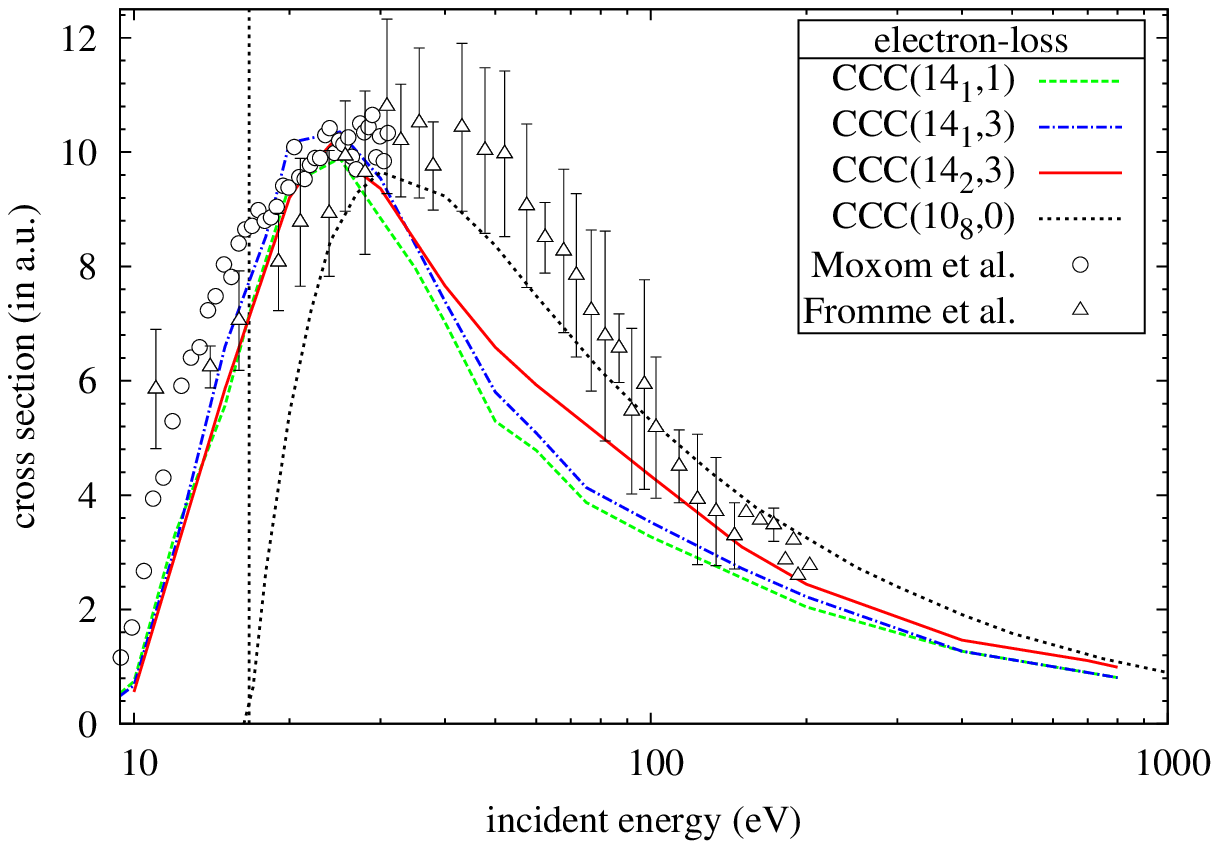}
\caption{The total electron-loss cross section for $e^+-$H$_2$ collisions. Experimental data are due to
\citet{MLC93} and \citet{FKRS88}.
The CCC calculations are due to  \citet{ZFB13r} (one-centre) and
\citet{Uetal15} (two-centre).
\label{Figure6H2}
}
\end{center}
\end{figure}

In Fig.~\ref{Figure5H2} the CCC results for the direct total ionization cross section (TICS) are compared with the available experimental data.
The experimental data of \citet{FKRS88} and \citet{KBCP90}
are in agreement with each other but differ from measurements of
\citet{JFKM95}  between 30 and 100~eV.  The largest CCC calculation
CCC(14$_2$,3), which contains $s$-, $p$- and $d$-atomic orbitals
together with the three lowest-energy Ps states, is in better agreement
with the measurements of \citet{JFKM95}. The 
CCC(14$_1$,1) and CCC(14$_1$,3) are systematically lower, primarily
due to the absence of the $d$-atomic orbitals.  Due to the unitarity of the close-coupling
formalism, larger $l$-orbitals
are likely to only marginally increase the TICS. Once we have convergence in the elastic scattering (large
$l$-orbitals not required) the GTCS is set, with the distribution
between elastic, electron-excitation, Ps-formation and total ionisation
cross sections thereby being constrained~\cite{B94l}.

Finally, the total electron-loss cross section is given in
Fig.~\ref{Figure6H2}, and is the sum of TICS and Ps-formation cross
sections. It is useful because the one-centre CCC approach is able to
yield this cross section at energies above the ionisation
threshold. This provides for an important internal-consistency
check. We see that around 50~eV the largest two-centre CCC calculation
is somewhat lower than the one-centre calculation. Given the increase
in the cross section due to the inclusion of $d$-orbitals adding
$f$-orbitals would go someway to reduce the discrepancy. Increasing
the Laguerre basis $N_l=10-l$ in the one-centre calculations would
increase the cross section just above the ionisation threshold due to
an improving discretisation of the continuum. 

Despite the immense complexity of the positron-molecular-hydrogen
scattering problem considerable progress in obtaining reasonably
accurate cross sections has been made.

\section{Antihydrogen formation}
\label{antih}
As stated earlier, via Eq.~\eref{antiHform}, antihydrogen formation
is effectively the time-reverse process of Ps formation upon
positron-hydrogen scattering. Accordingly, it only takes place at
positron energies above the Ps($n$) formation threshold, where $n$
indicates the principal quantum number. Furthermore, we require
two-centre calculations because only these have explicit Ps and
H states. Given that we are interested in as large cross sections as
possible, and that the exothermic transition cross sections go to infinity at
threshold~\cite{wigner_behavior_1948,F74} the primary energy range of
interest is within the extended Ore gap, as given in \fref{p-H}.

\begin{figure}[htb]
\includegraphics[width=\columnwidth]{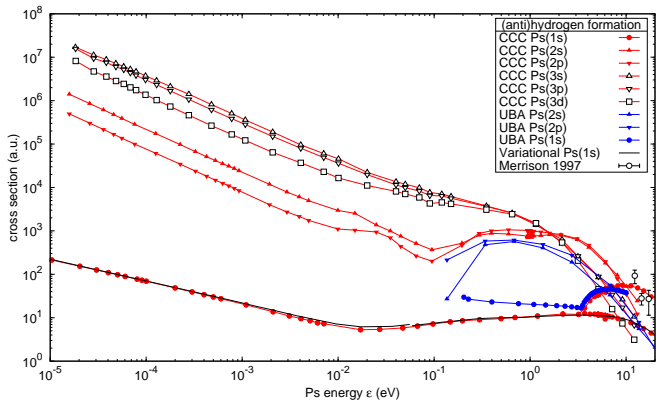}
\caption{Total cross sections for (anti)hydrogen formation upon positronium, in the specified initial state $nl$, scattering on
(anti)protons calculated using the CCC
  method, first presented by \citet{Ketal15l}.  For Ps(1s), the variational calculations
\cite{Humberston87,Hetal97,HR15} are for (anti)hydrogen formation in the 1s state
only (CCC-calculated unconnected points presented for comparison),
while the UBA calculations of
\citet{mitroy_formation_1995} and \citet{MS94}, and the CCC calculations generally,
are for (anti)hydrogen formation in all open states. The three
experimental points are due to \citet{Merrison97}.}
\label{p-Ps}
\end{figure}

There are few calculations of antihydrogen formation that are accurate
at low energies. As far as we are aware, apart from the CCC
method~\cite{Ketal15l,Retal16}, 
only the variational approach of \citet{Humberston87} has yielded
accurate results. The
unitarised Born approximation (UBA)
calculations~\cite{Mitroy95b,Mitroy95,mitroy_formation_1995} are not
appropriate at low energies, and neither are first order
approximations~\cite{CH13}.

Large cross sections for antihydrogen formation
occur for transitions between excited states of Ps 
and H. To have these states accurate the Laguerre bases $N_l$ for both H and
Ps need to be sufficiently large. \citet{Ketal15l} used $N_l=12-l$,
with the resultant energies given in \fref{energies}. While it is
trivial to have larger $N_l$ as far as the structure is concerned, the
primary limiting factor is to be able to solve the resultant
close-coupling equations. In \fref{energies} the positive energies
correspond to the breakup of H and also of Ps, and yet the two
represent the same physical process. This leads to highly
ill-conditioned equations. \citet{BKB15} considered non-symmetric
treatments of the two centres by dropping the Ps positive-energy
states. While these also satisfy internal-consistency checks they have
not proved to be more efficient in yielding convergence, with
increasing $N_l$, in the cross
sections of present interest. 

\subsection{Numerical
  treatment of the Green's function}
A comprehensive set of antihydrogen formation cross sections for low-energy
Ps incident on an antiproton has been given by \citet{Retal16}. They
used the original numerical treatment of the
Green's function in Eq.~\eref{LSKmat}. 
In \fref{p-Ps}, we present the summary of the total antihydrogen
formation cross sections for specified initial Ps($nl$) as
presented by \citet{Ketal15l}. Excellent agreement
with the variational calculations~\cite{Humberston87,Hetal97,HR15},
available only for the ground states of H and Ps, gives us confidence in
the rest of the presented CCC results. 

As predicted by \citet{wigner_behavior_1948} the ground state cross
section goes to infinity as $1/\sqrt{\varepsilon}$ at threshold, whereas for
excited states this is modified to $1/\varepsilon$ at threshold due to
the long-range dipole interaction of degenerate Ps $n\ge2$ states,  as
explained by \citet{F74}. The cross sections increase steadily with
increasing principle quantum number of Ps, with the transition to the
highest available principal quantum number of H being the most
  dominant contribution~\cite{Retal16}. Consequently there is
  considerable motivation to push the calculations even further to
  larger $N_l$.

\subsection{Analytic
treatment of the Green's function}
\label{analyticG}
There is a second source of ill-conditioning of the close-coupling
equations to that discussed previously. Given that we are particularly
interested in low energies i.e., just above the various thresholds,
the singularity in the Green's function of Eq.~\eref{LSKmat} occurs very
close to zero energy. A typically used symmetric treatment on either
side of the singularity results in very large positive and negative
values which contribute to the ill-conditioning via precision loss.  

	\begin{figure}[h]
		\centering
		\includegraphics[width=\columnwidth]{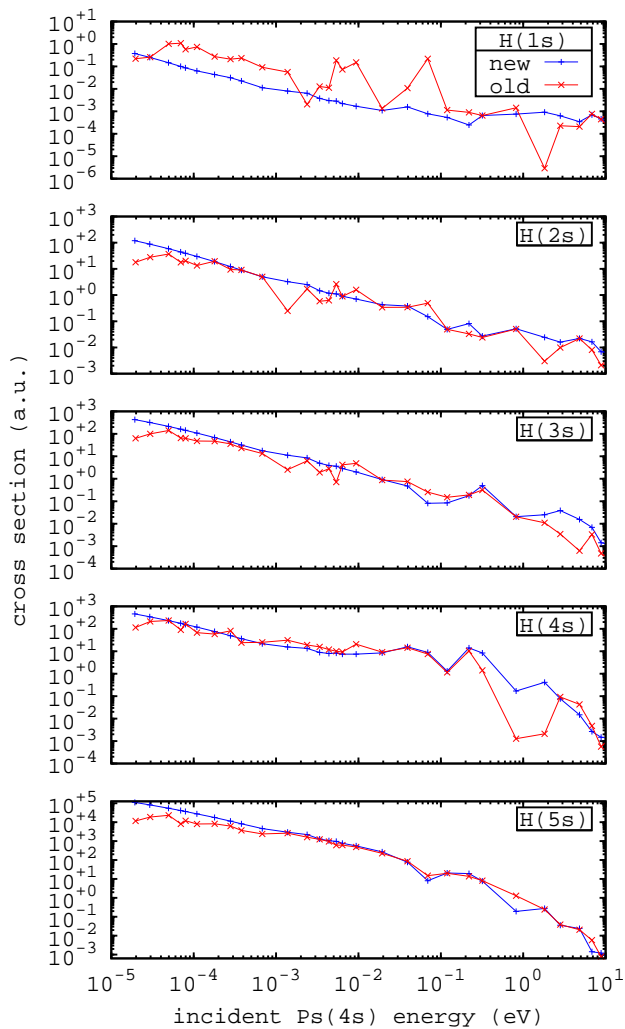}
		\caption{H($ns$) formation cross sections for $n\le5$
                  in p-Ps($4s$) scattering for the zeroth partial
                  wave. The ``old'' results 
                  arise from solving the original CCC
                  equations~\eref{LSKmat}. The ``new'' results are due
                  to the solution with the Green's function being
                  treated analytically, see \citet{Bcpc16}.}
		\label{new}
	\end{figure}

Very recently \citet{Bcpc15,Bcpc16} showed that the Green's
function of the coupled Lippmann-Schwinger equations~\eref{LSKmat} can
be treated analytically. The method that was previously used in calculating the
optical potential in the coupled-channel optical
method~\cite{BKM91e,BM93} was applied directly to the Green's function
in the Lippmann-Schwinger equation \eref{LSKmat}.
In doing so, they showed that they could improve
on the above $N_l=12-l$ basis to reach $N_l=15-l$, and thereby
consider transitions from Ps($n=4$) states. In \fref{new} we compare
the ``old'' and the ``new'' numerical formulations and find that the
latter is quite superior. In fact, for  $N_l=15-l$ \citet{Bcpc16} were unable to
obtain numerical stability in the original formulation with variation
of integration parameters 
over the momentum in Eq.~\eref{LSKmat}, and presented
just a combination that yielded somewhat reasonable agreement with
the results of the ``new'' numerical method. The new technique also
proved to be particularly advantageous in studying threshold phenomena
in positronium-(anti)proton scattering~\cite{FBKB16}.

\section{Concluding remarks}

We have presented an overview of recent developments in
positron-scattering theory on several atoms and molecular hydrogen,
with particular emphasis on two-centre calculations that are able to
explicitly treat positronium formation. While considerable progress
has been made there remain some major discrepancies between theory and
experiment such as for low-energy positron-sodium
scattering. Considerable technical development is still required for
complicated atoms such as the heavier inert gases to incorporate
Ps formation in a systematically convergent way. A general scheme for
doing so for molecular targets is also required.

It is also important to state that there are still fundamental issues to be
addressed in the case of breakup. While no overcompleteness has been
found in two-centre calculations with near-complete bases on both
centres, determining the resulting differential 
cross sections remains an unsolved problem. 
\citet{KBBS14} considered the simplest energy-differential cross
section which describes the probablity of an electron of a certain
energy being ejected. If we have contributions to this process from
both centres, how do we combine them? Furthermore, diagonalising
the Ps Hamiltonian yields pseudostates of
positive-energy Ps rather than the energy of the individual electron or positron. It may be that the only practical way
to obtain differential ionisation cross sections is to restrict the Ps
centre to just negative-energy eigenstates, and obtain differential
cross sections solely from the atomic centre positive-energy
pseudostates, as is done for electron-impact
ionisation~\cite{Bpr12}. However, this kind of approach may not be
capable of describing the phenomenon of Ps formation in continuum as seen by \citet{AKL05}.
This is currently under investigation.

\begin{acknowledgments}
Support of the Australian Research Council and the Merit 
Allocation Scheme on the National Facility of the Australian
Partnership for Advanced Computing is gratefully acknowledged. This
work was supported by resources provided by Pawsey Supercomputing
Centre with funding from the Australian Government and the Government
of Western Australia. A.S.K. acknowledges partial support from the
U.S. National Science Foundation under Award No. PHY-1415656. 
\end{acknowledgments}


\end{document}